\documentclass[twoside,twocolumn,10pt]{elsartjc}
\usepackage{times}
\usepackage{amsmath}
\usepackage{graphicx}
\usepackage{amssymb}

\begin{document}

\def\d{\partial}
\def\um{\,\mu{\rm  m}}
\def\mm{\,   {\rm mm}}
\def\cm{\,   {\rm cm}}
\def \m{\,   {\rm  m}}
\def\ps{\,   {\rm ps}}
\def\ns{\,   {\rm ns}}
\def\us{\,\mu{\rm  s}}
\def\ms{\,   {\rm ms}}
\def\nA{\,   {\rm nA}}
\def\uA{\,\mu{\rm  A}}
\def\mA{\,   {\rm mA}}
\def\A {\,   {\rm  A}}
\def\mV{\,   {\rm mV}}
\def\V {\,   {\rm  V}}
\def\fF{\,   {\rm fF}}
\def\pF{\,   {\rm pF}}
\def\GeV{\, {\rm GeV}}
\def\MHz{\, {\rm MHz}}
\def\uW{\,\mu{\rm  W}}
\def\e {\,  {\rm e^-}}

\renewcommand{\labelenumi}{\arabic{enumi}}
\renewcommand{\labelitemi}{-}

\begin{frontmatter}

\title{Pixel Detectors}
\author {N.~Wermes\thanksref{BMBF} }

\thanks[BMBF]{Work supported by the German Ministerium f{\"u}r Bildung,
              und Forschung (BMBF) under contract
              no.~$05 HA1PD1/5$\ and
              by the DIP Foundation under contract no. E7.1\\
              address: Physikalisches Institut, Nussallee 12,
               D-53115 Bonn, Germany, Tel.: +49\,228\,73-3533, Fax:
               -3220, email: wermes@uni-bonn.de}

\address{Physikalisches Institut der Universit{\"a}t Bonn, Germany}

%
%
%
\end{frontmatter}

%
\large {Abstract}\par
Pixel detectors for precise particle tracking in high energy
physics have been developed to a level of maturity during the past
decade. Three of the LHC detectors will use vertex detectors close
to the interaction point based on the hybrid pixel technology
which can be considered the ´`state of the art´` in this field of
instrumentation. A development period of almost 10 years has
resulted in pixel detector modules which can stand the extreme
rate and timing requirements as well as the very harsh radiation
environment at the LHC without severe compromises in performance.
From these developments a number of different applications have
spun off, most notably for biomedical imaging. Beyond hybrid
pixels, a number of monolithic or semi-monolithic developments,
which do not require complicated hybridization but come as single
sensor/IC entities, have appeared and are currently developed to
greater maturity. Most advanced in terms of maturity are so called
CMOS active pixels and DEPFET pixels. The present state in the
construction of the hybrid pixel detectors for the LHC experiments
together with some hybrid pixel detector spin-off is reviewed. In
addition, new developments in monolithic or semi-monolithic pixel
devices are summarized.

\section{Hybrid pixel detectors for the LHC experiments}
The truly challenging requirements on detectors operation close to
the interaction points at the LHC are on spatial resolution, on
timing precision, and most importantly on the long term operation
performance and radiation tolerance to particle fluences as high
as $10^{15}$n$_{eq}/$cm$^{-2}$. At present, these demands are only
met by so-called hybrid pixel detectors, for which the particle
sensing element, the sensor, and the integrated electronics
circuitry, the readout chip, are separate entities. They are mated
by a hybridization technique, known as bump and flip-chip
technology. All of the LHC-collider-detectors ALICE
\cite{ALICE_pix,ALICE_Riedler}, ATLAS
\cite{ATLAS_pix,ATLAS_Gemme}, and CMS \cite{CMS_pix,CMS_Erdmann},
LHCb (for the RICH system) \cite{LHCb_pix} at the LHC, and the
CERN fixed target experiment NA60~\cite{NA60}, employ the hybrid
pixel technique to build large scale (up to $\sim$2m$^2$) pixel
detectors. Pixel area sizes are typically 50 $\times$ 400
$\mu$m$^2$ as for ATLAS or 100 $\times$ 150 $\mu$m$^2$ as for CMS.
The detectors are arranged in cylindrical barrels of 2 to 3 layers
and disks covering the forward and backward regions.

\subsection{The sensors}
\begin{figure}[b]
\begin{center}
\includegraphics[width=0.42\textwidth]{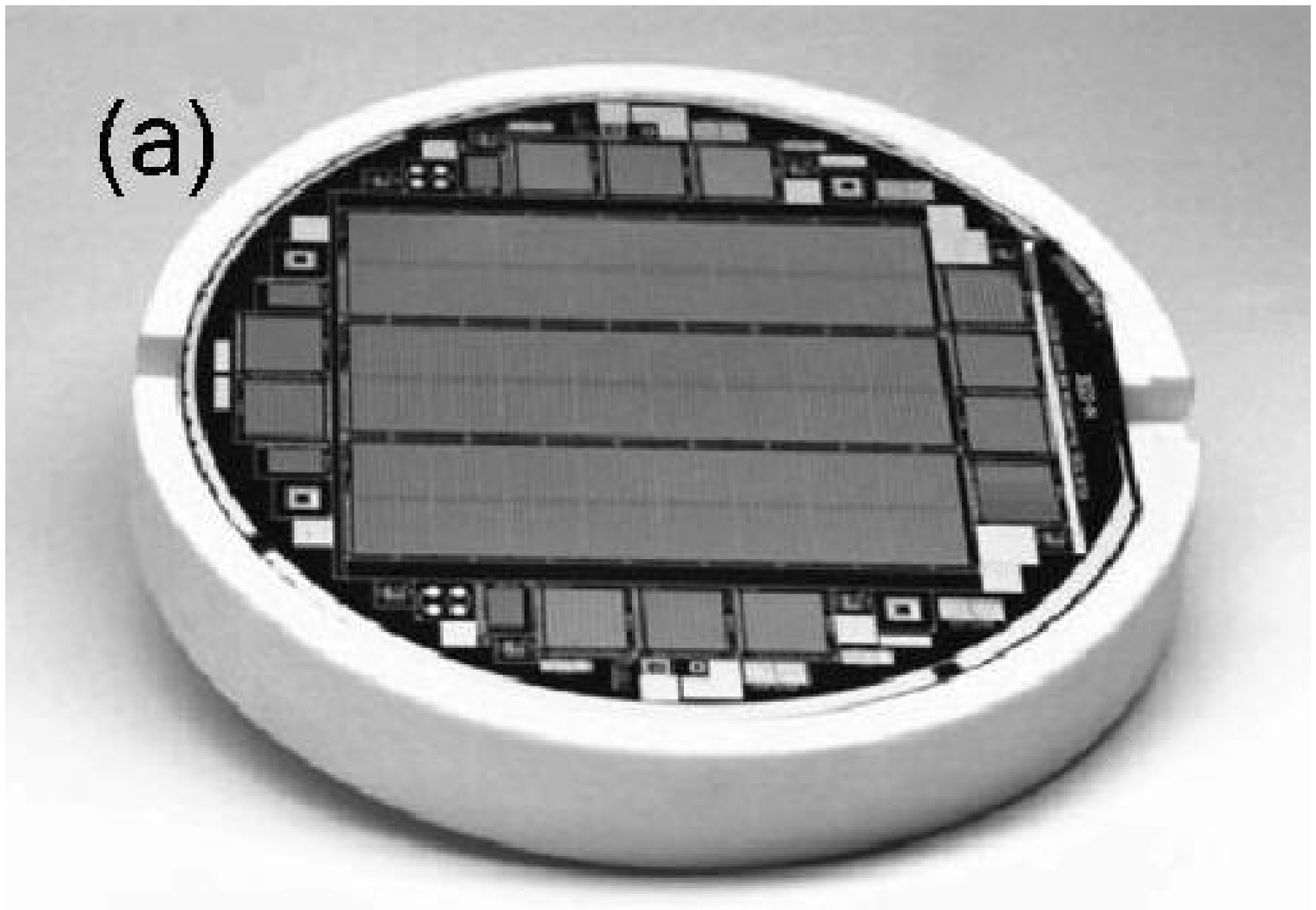}
\hspace{0.1cm}
\includegraphics[width=0.42\textwidth]{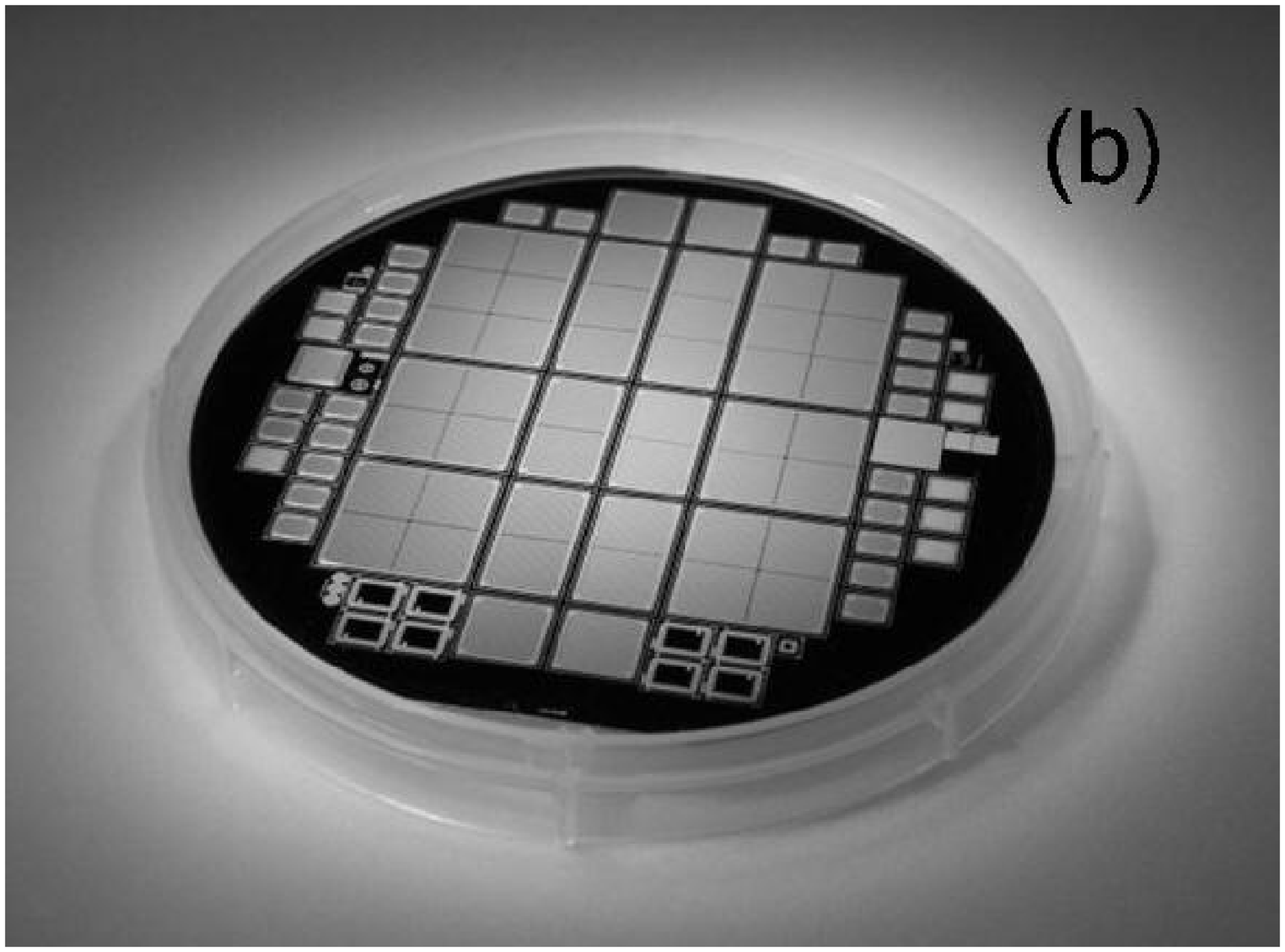}
\end{center}
\caption[]{Photographs of the ATLAS (a) and CMS-disks (b) pixel
sensor wafers.} \label{ATLAS-Sensor}
\end{figure}

The discovery that oxygenated silicon is more radiation hard, with
respect to the non-ionizing energy loss of protons and
pions~\cite{oxysilicon} than standard silicon, allows operation of
pixel detectors at the LHC for which the radiation is most severe
due to their proximity to the interaction point. Sensors with
$n^+$ electrodes in n-bulk material have been chosen to cope with
the fact that type inversion occurs after about $\Phi_{eq} = 2
\times 10^{12}$cm$^{-2}$. After type inversion to p-type bulk
material the $pn$-diode sits on the electrode side, from which the
depletion zone develops into the bulk, thus allowing the sensor to
be operated partially depleted. For the Super-LHC, a name termed
for an LHC-upgrade programme which targets a luminosity of
10$^{35}$cm$^{-2}$s$^{-1}$ and hence scales the environment at the
LHC by a factor ten in all aspects, new sensor technologies are
needed to cope with the radiation hardness demands. Figures
\ref{ATLAS-Sensor}(a) and (b) show photographs of the sensor
wafers of ATLAS and CMS, respectively, which both use oxygenated
silicon as the sensor material.

\subsection{The FE electronics}
\begin{figure}[b]
\begin{center}
\hspace{0.6cm}
\includegraphics[width=0.35\textwidth]{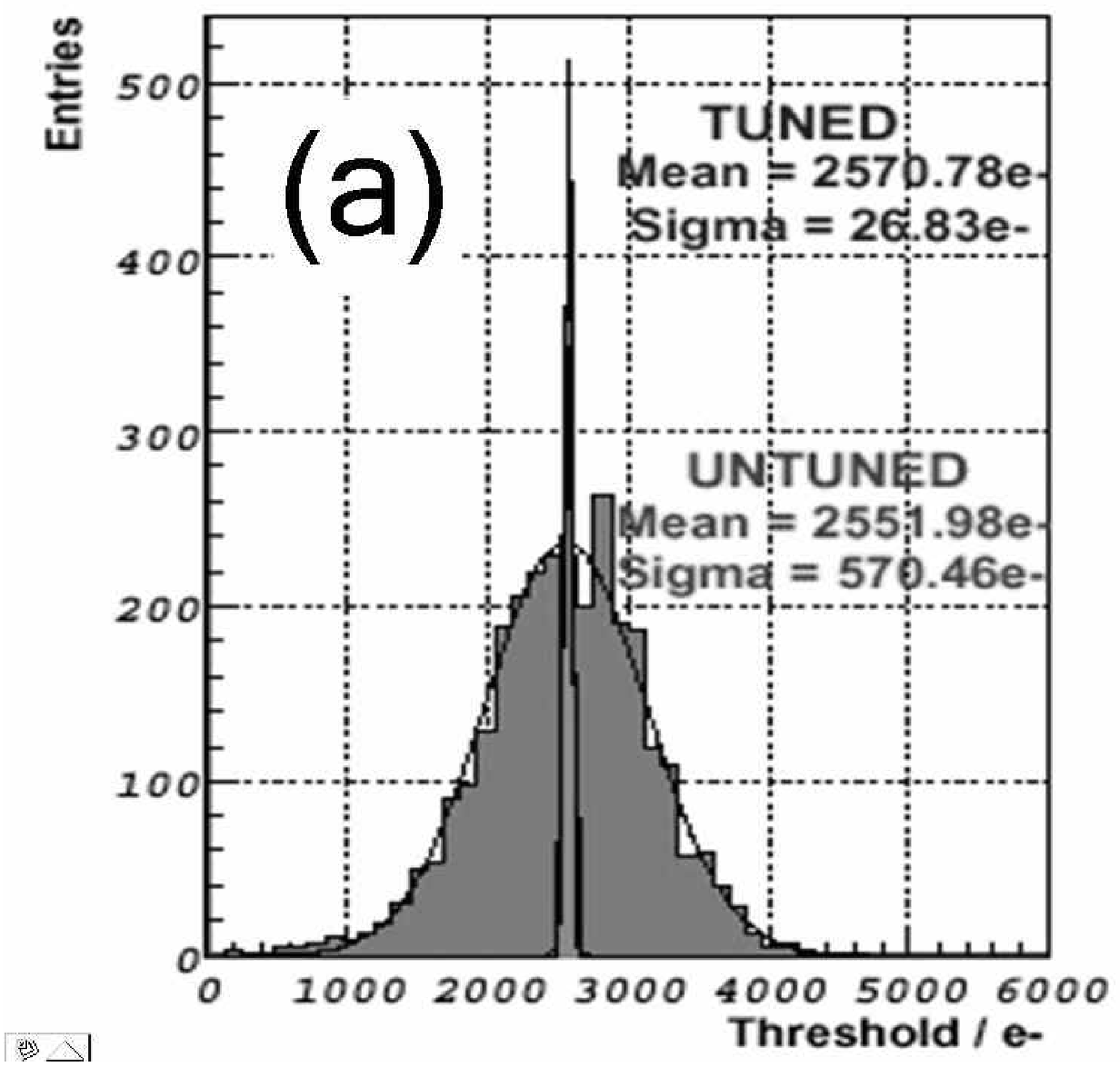}
\hfill \vspace{0.1cm} \break
\includegraphics[width=0.48\textwidth]{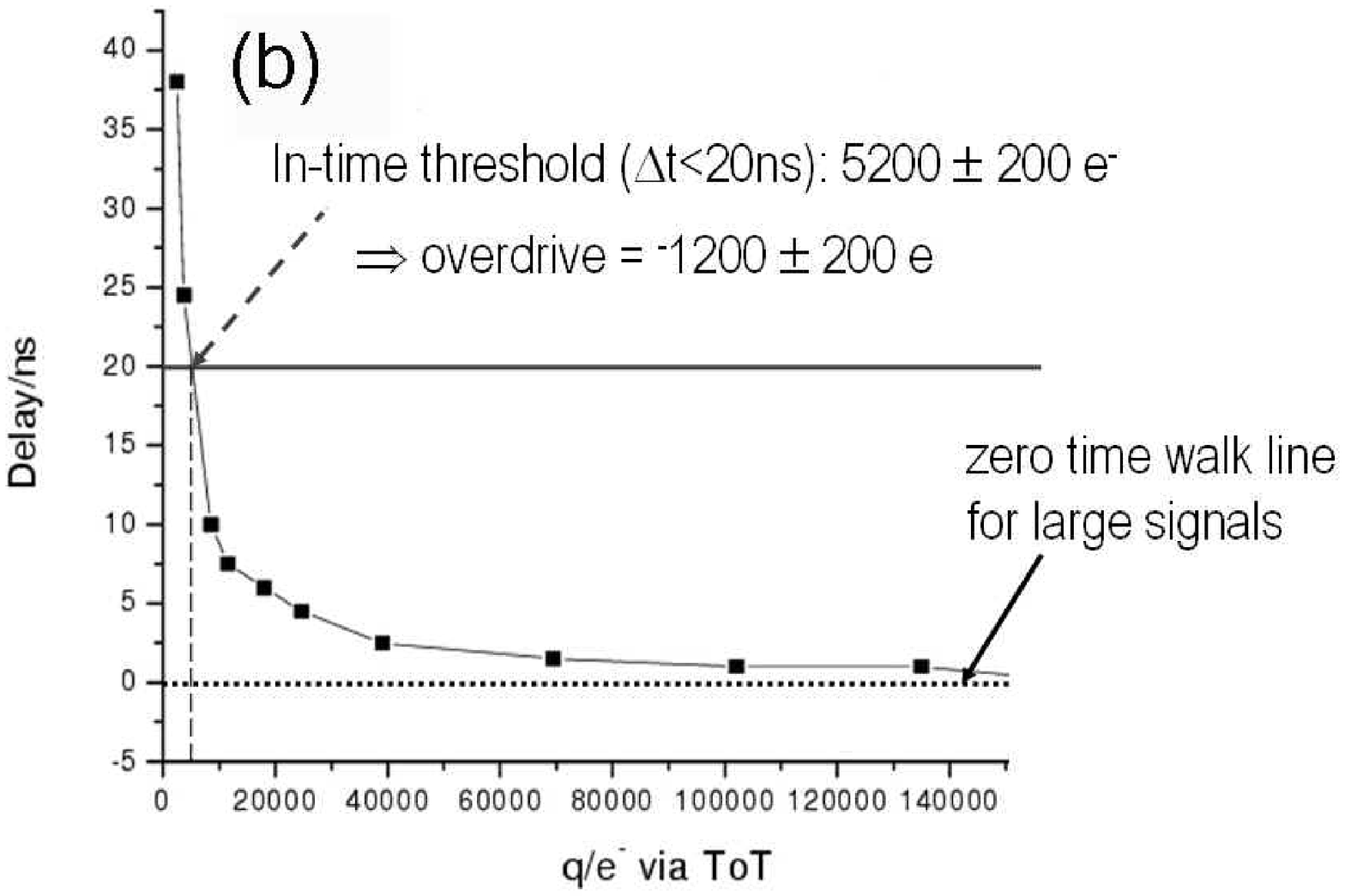}
\end{center}
\caption[]{(a) Dispersion of the pixel thresholds before and after
tuning. (b) In-time threshold and overdrive for a typical
threshold setting of 4000$\pm$200$e^-$.} \label{thresholds}
\end{figure}
The challenge in the design of the front-end pixel electronics
\cite{blanquart03} can be summarized by the following
requirements: low power ($\lesssim$ 50$\mu$W per pixel), low noise
and threshold dispersion (together $\lesssim$ 200e), zero
suppression in every pixel, on-chip hit buffering, and small
time-walk to be able to assign the hits to their respective LHC
bunch crossing. The pixel groups at the LHC have reached these
goals in several design iterations using first
radiation-\emph{soft} prototypes, then dedicated radhard designs,
and finally using deep submicron technologies. For ATLAS the full
production quantity of chip wafers has been processed and tested
with an average yield of 82$\%$. CMS yields are in the same order
and ALICE chip yields are 51$\%$ with a chip area of 13.5 $\times$
15.8 mm$^2$. While CMS uses analog readout of hits, ATLAS obtains
pulse height information by means of measuring the \emph{time over
threshold} (ToT) for every hit. Figure \ref{thresholds}(a) shows
the distribution of measured thresholds of an ATLAS front-end
chip. The dispersion of about 600 e$^-$ can be lowered to below 50
e$^-$ by a 7-bit tuning feature implemented in the chip. Figure
\ref{thresholds}(b) illustrates the effect of time walk for small
signals. For efficient signal detection within a defined time of
20 ns with respect to the bunch crossing an {\it overdrive} of
about 1200$e^-$ is necessary. The bunch crossing occurs every 25
ns.

The hit information is extracted as follows: A 40 MHz Gray coded
clock is transmitted to all pixel cells. If the pixel circuit
detects a hit signal (analog) it generates digital hit
information. The hit data (address and time stamp) are transmitted
to the bottom of the chip and temporarily stored in end of column
buffers outside pixel matrix. The buffers monitor the age of each
hit data and delete hits when no trigger coincidence occurs. Hits
having their time stamp coincident with the LV1 trigger are
finally read out.

\subsection{Hybridization}
\begin{figure}[h]
\includegraphics[width=0.3\textwidth]{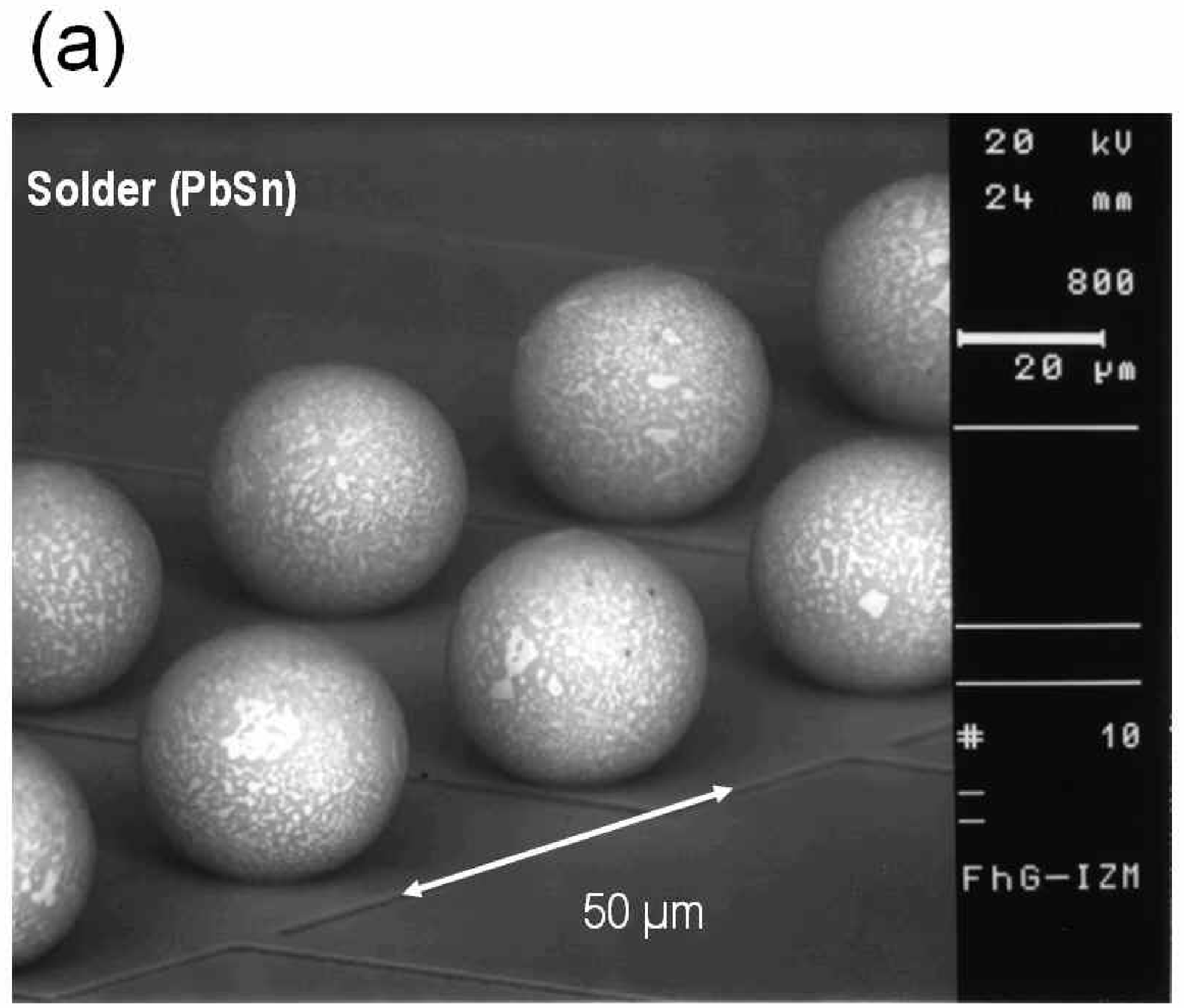}
\includegraphics[width=0.33\textwidth]{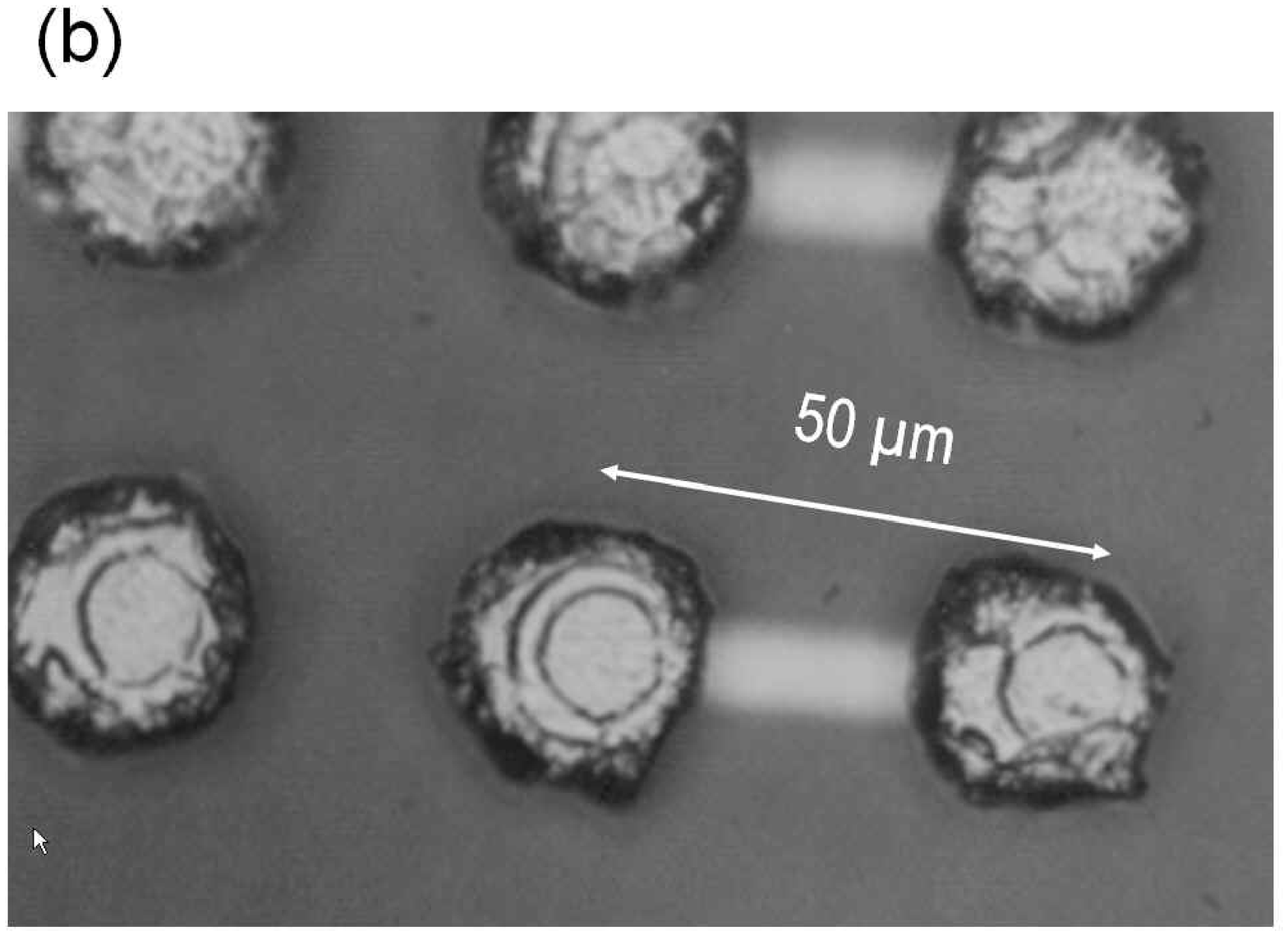}
\includegraphics[width=0.15\textwidth]{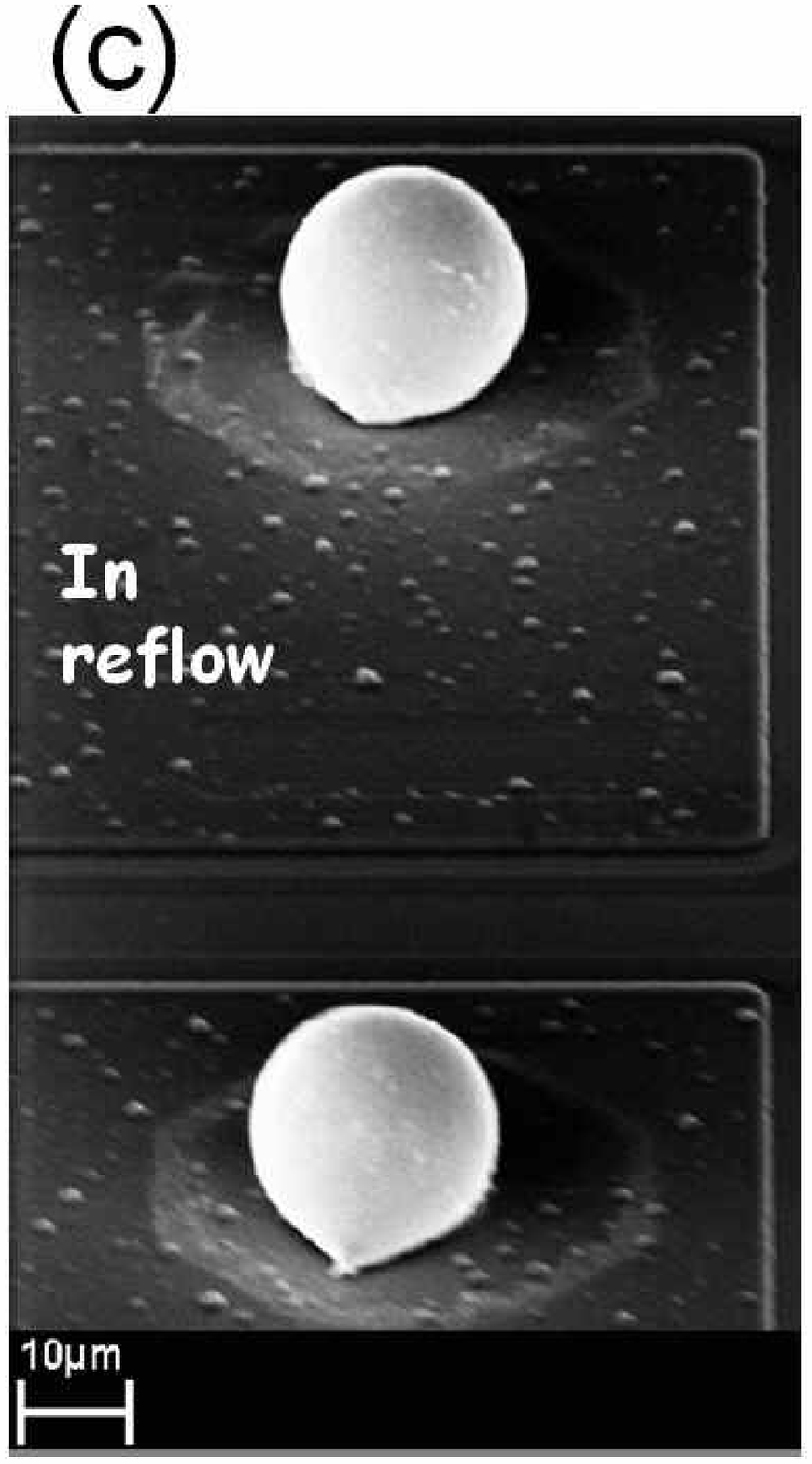}
\caption{(a) solder (PbSn, Photo IZM, Berlin) (b) Indium (Photo
AMS, Rome), and (c) Indium with reflow (Photo PSI, Villigen) bump
rows with $50 \mu$m pitch.} \label{bumps}
\end{figure}
\begin{figure}[htb]
\includegraphics[width=.45\textwidth]{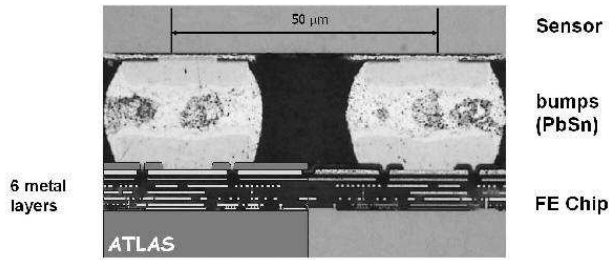}
\caption{SEM cross section (IZM, Berlin) of a bumped ATLAS
assembly with the sensor on the top and the FE-chip on the
bottom.} \label{bumps}
\end{figure}

Hybridization of chip and sensor is done by fine pitch bumping and
subsequent flip-chipping, either with PbSn (solder) or with Indium
bumps. The Indium bumps are applied by a wet lift-off technique
and can be mated by direct thermo-compression
\cite{Indium_bumping,fiorello} or reflown, as developed by CMS
\cite{pixbook}. After bumping the chips are thinned by backside
grinding to a thickness of 150 - 180 $\mu$m. Fig. \ref{bumps}
shows rows of $50 \mu$m pitch bumps obtained by these techniques.
All of these bump bonding technologies have been successfully used
with 8" IC-wafers and 4" sensor wafers.

ATLAS pixel modules are hybridized by two vendors (one using
indium one solder) at a rate of about up to 2 $\times$ 20 per
week. The fraction of broken, bridged or missing bumps is at the
level of 10$^{-4}$. Above 85$\%$ of all modules have less than
0.1$\%$ of bad bumps. About 10-15$\%$ of the modules arriving from
the vendor have a chip with an unacceptably high number of bump
failures and need to be reworked, i.e. removing the FE-chip from
the module and flip-chipping of a new one. The success fraction of
this operation is 99$\%$ for solder and 80$\%$ for indium, mostly
due to the fact that In-bumps are flatter and remnants of dirt in
the reworking process do more harm. The total reject fraction of
modules to date is 1$\%$ and 14$\%$~\cite{attilio} for PbSn- and
Indium bumped modules, respectively.

\subsection{The modules}
\begin{figure}[htb]
\hspace{0.5cm}\includegraphics[width=0.42\textwidth]{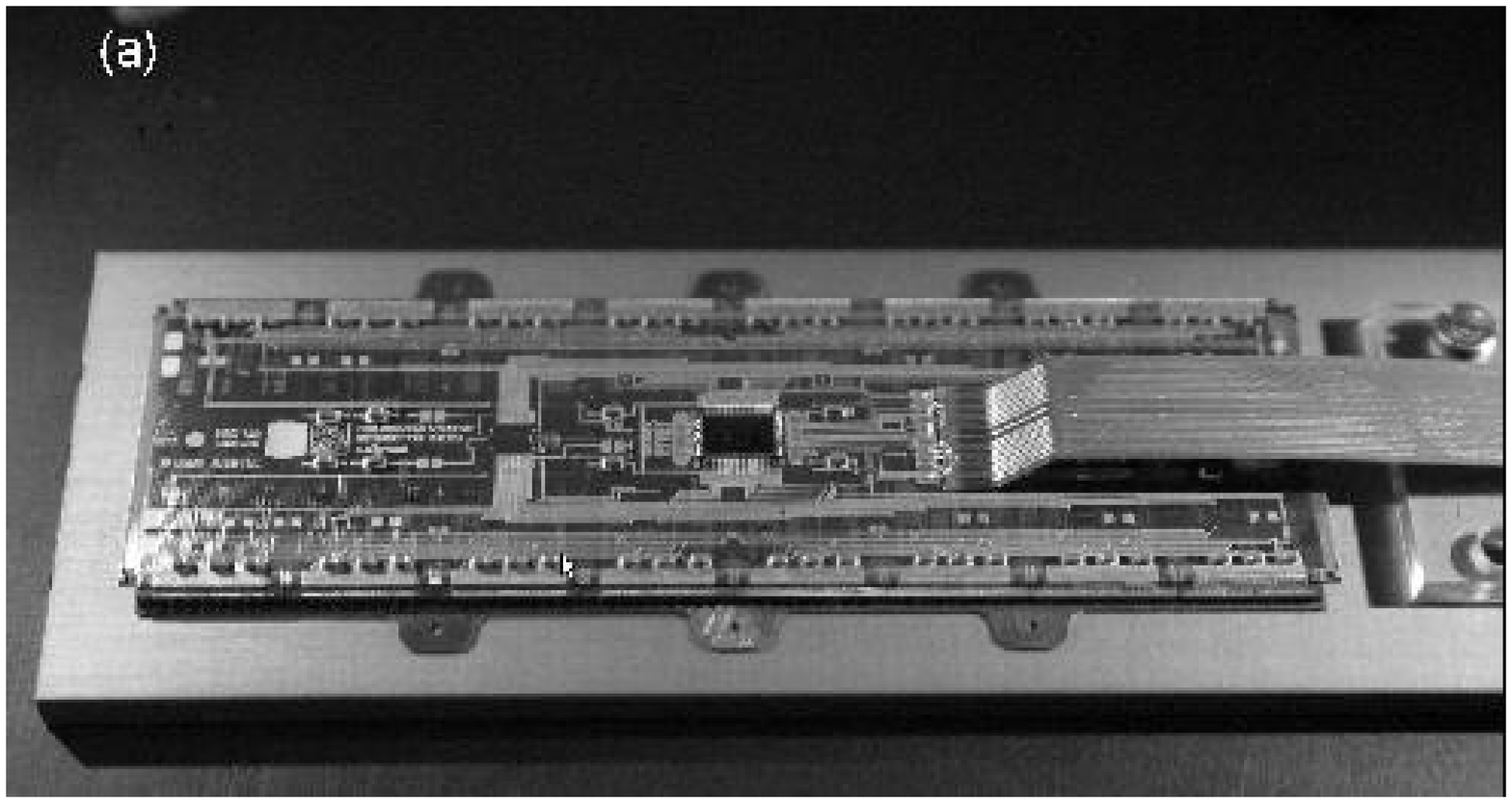}
\hfill \vspace{0.2cm} \break
\includegraphics[width=0.48\textwidth]{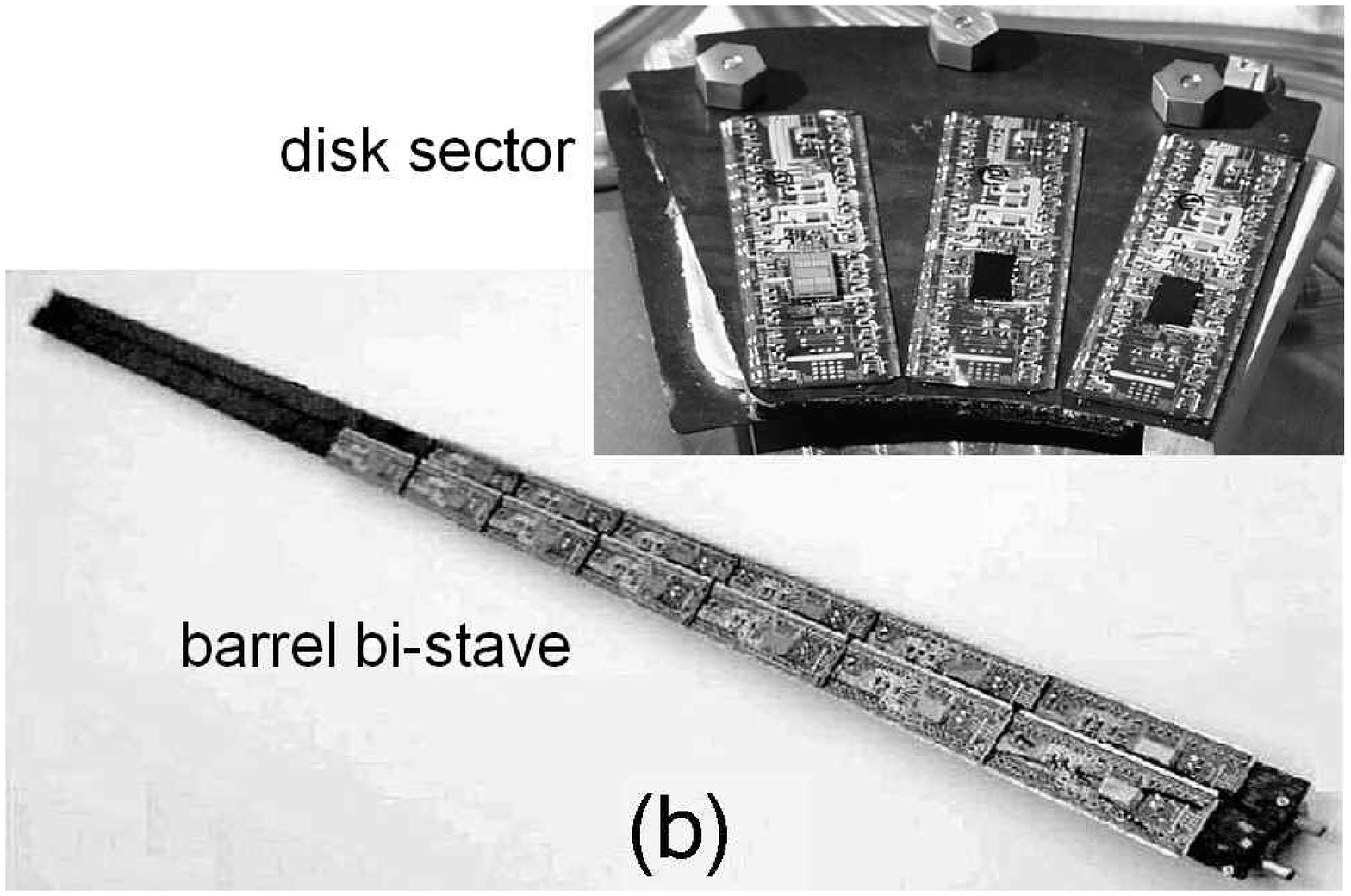}
  \caption{(a) Assembled CMS-pixel module with one sensor and 16 readout chips.
  (b) ATLAS modules mounted to a bi-stave unit and to a disk sector.}
  \label{modules}
\end{figure}

The CMS and ATLAS modules (cf Fig.~\ref{modules}(a)) typically
have 2 cm $\times$ 6.5 cm area consisting of 16 FE-chips
bump-connected to one silicon sensor.The I/O lines of the chips
are connected via wire bonds to a kapton flex circuit glued atop
the sensor. The flex houses a module control chip responsible for
front end time/trigger control as well as event building. The
total thickness at normal incidence is in excess of $2.5\%$ $X_0$.
The modules are arranged in barrel-ladders or disk-sectors as
shown in Fig.~\ref{modules}(b) for the case of ATLAS.

\begin{figure}[htb]
\includegraphics[width=0.5\textwidth]{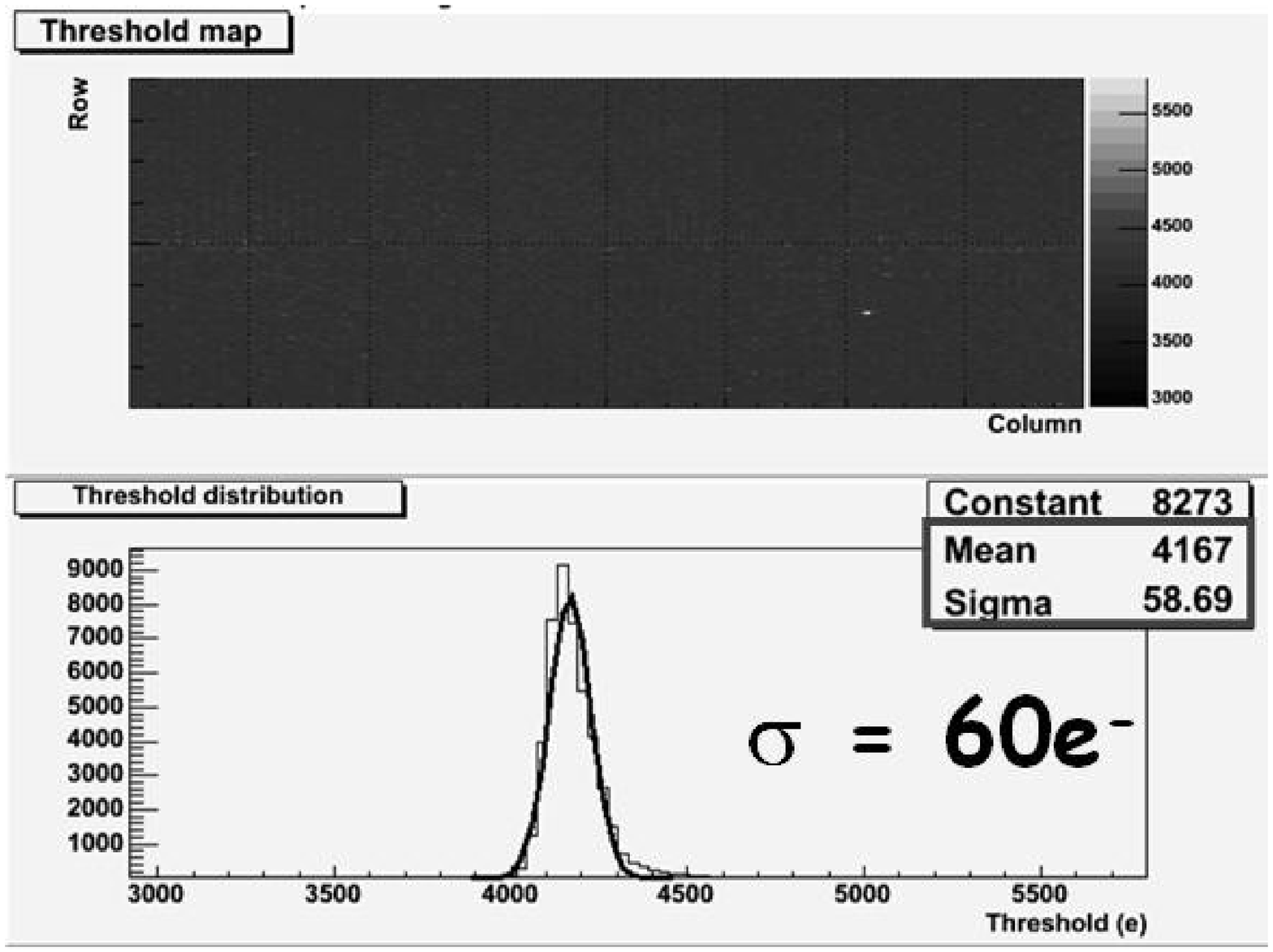}
\hfill \vspace{0.2cm} \break
\includegraphics[width=0.5\textwidth]{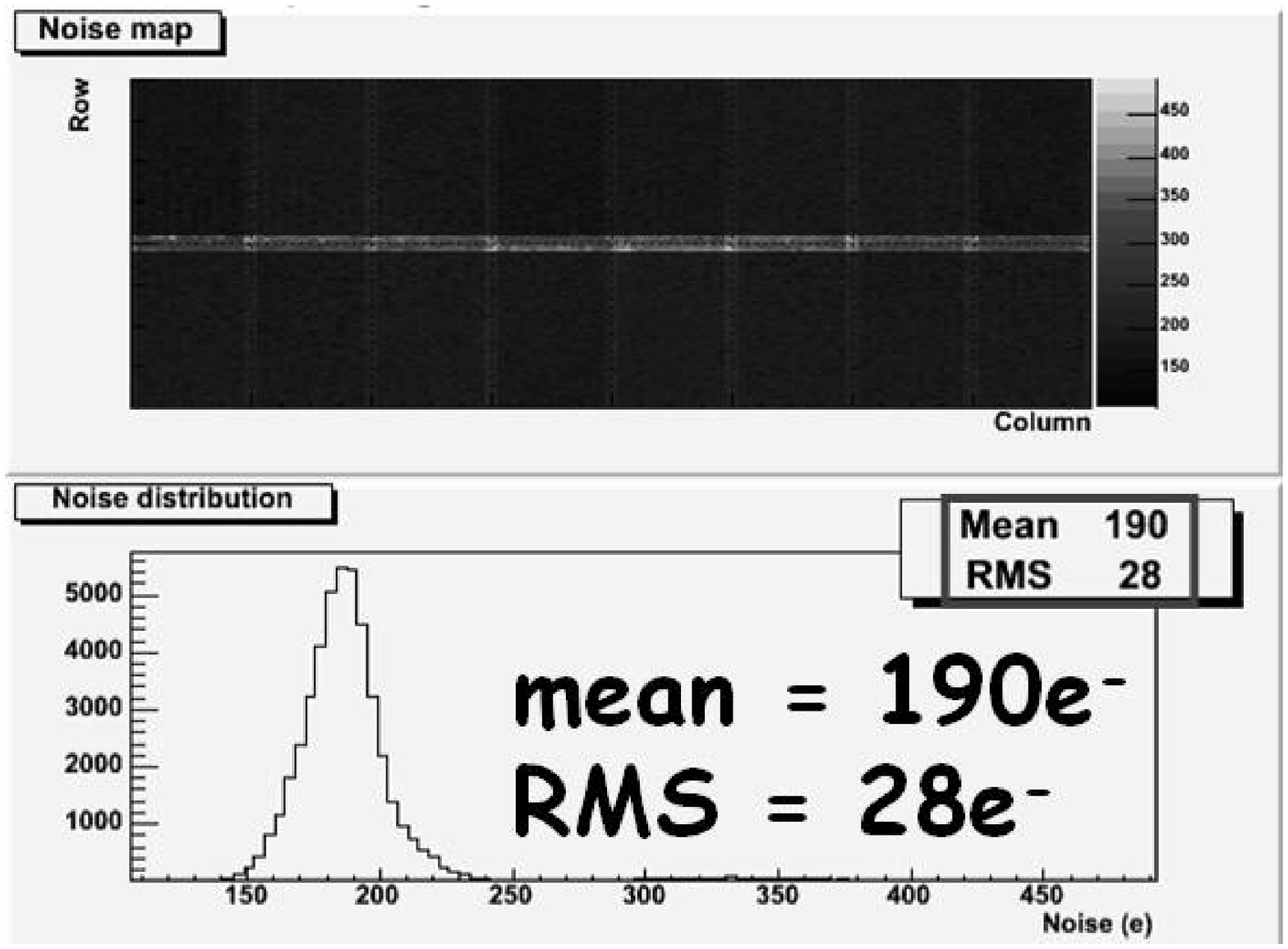}
  \caption{Threshold (top) and noise (bottom) distributions and
  maps of ATLAS production modules.}
 \label{noise}
\end{figure}

Production pixel modules are scrutinized by extensive tests in the
lab, among them their analog and digital functionality, noise and
threshold performance, response to an X-ray source scan, as well
as temperature cycling of modules and also of assembled ladders.
For ATLAS modules, after tuning the threshold spread is well below
100$e^-$ and the mean value of the noise distribution is between
150$e^-$ and 200$e^-$ (cf. Fig.~\ref{noise}). The quadratic sum of
both is many sigmas away from the typical threshold setting of
3000$e^-$ or above, a requirement at LHC in order to keep the
noise hit occupancy low. In order to qualify the modules for their
placement inside the ATLAS pixel detector arrangement, a ranking
factor based on pixel efficiency, sensor quality, noise and
threshold performance, and rework penalty is introduced and a cut
is placed to qualify the modules as barrel layer B,~1,~or~2 or as
disk modules. Based on a sample of about 1300 modules to date
54$\%$ of the produced modules meet the most demanding
qualification as B-layer modules. Finally, complete ladders of 13
module undergo a so-called system test, i.e. a test procedure
involving the complete readout chain of ATLAS, including
micro-cables, conversion into optical signal and clock
transmission and their routing over the full cable length in
ATLAS. Comparison to module stand-alone tests show no significant
chances in the performance.
\subsection{Radiation tolerance}
\begin{figure}[htb]
\begin{center}
\includegraphics[width=0.22\textwidth]{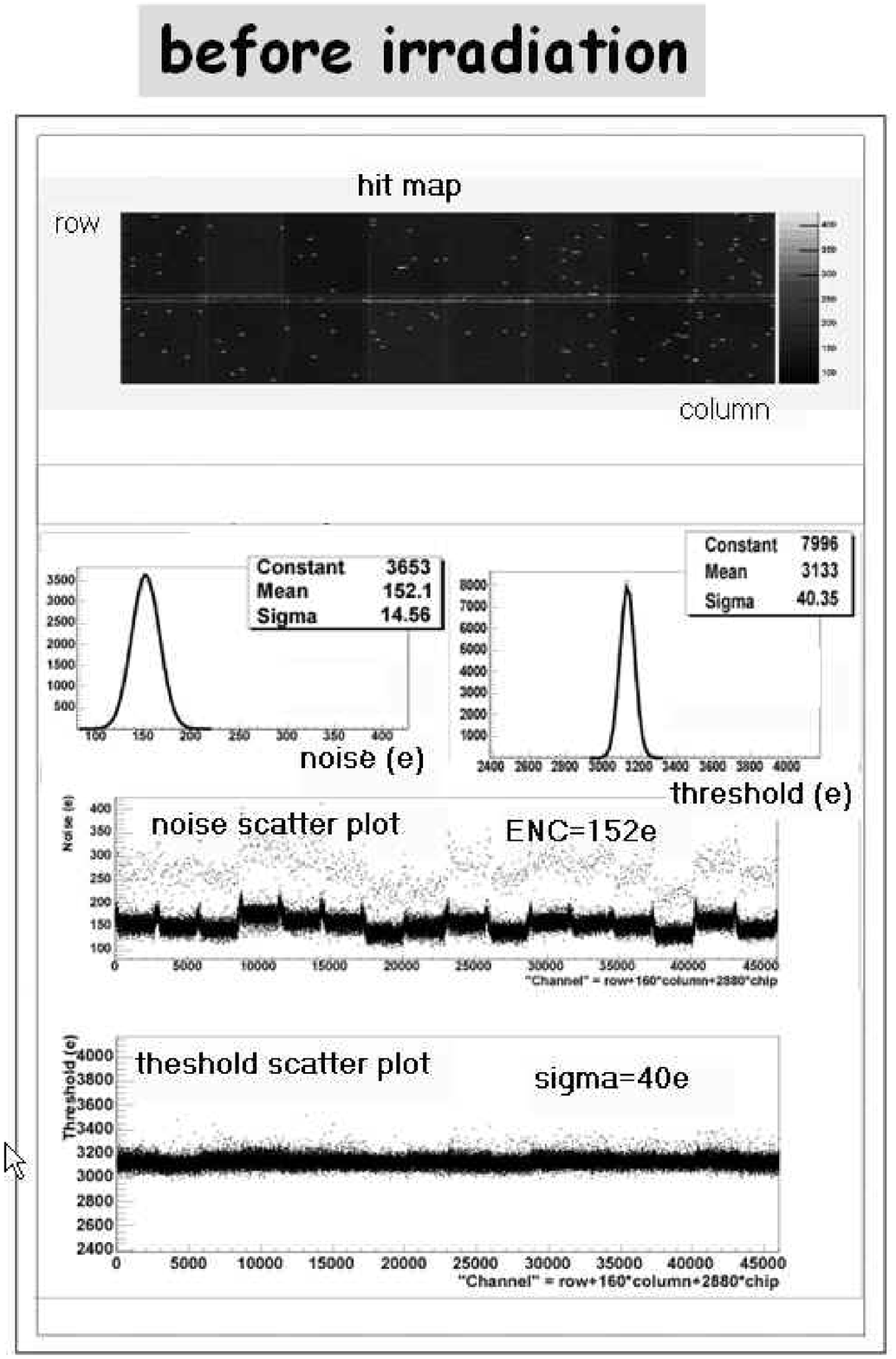}
\includegraphics[width=0.22\textwidth]{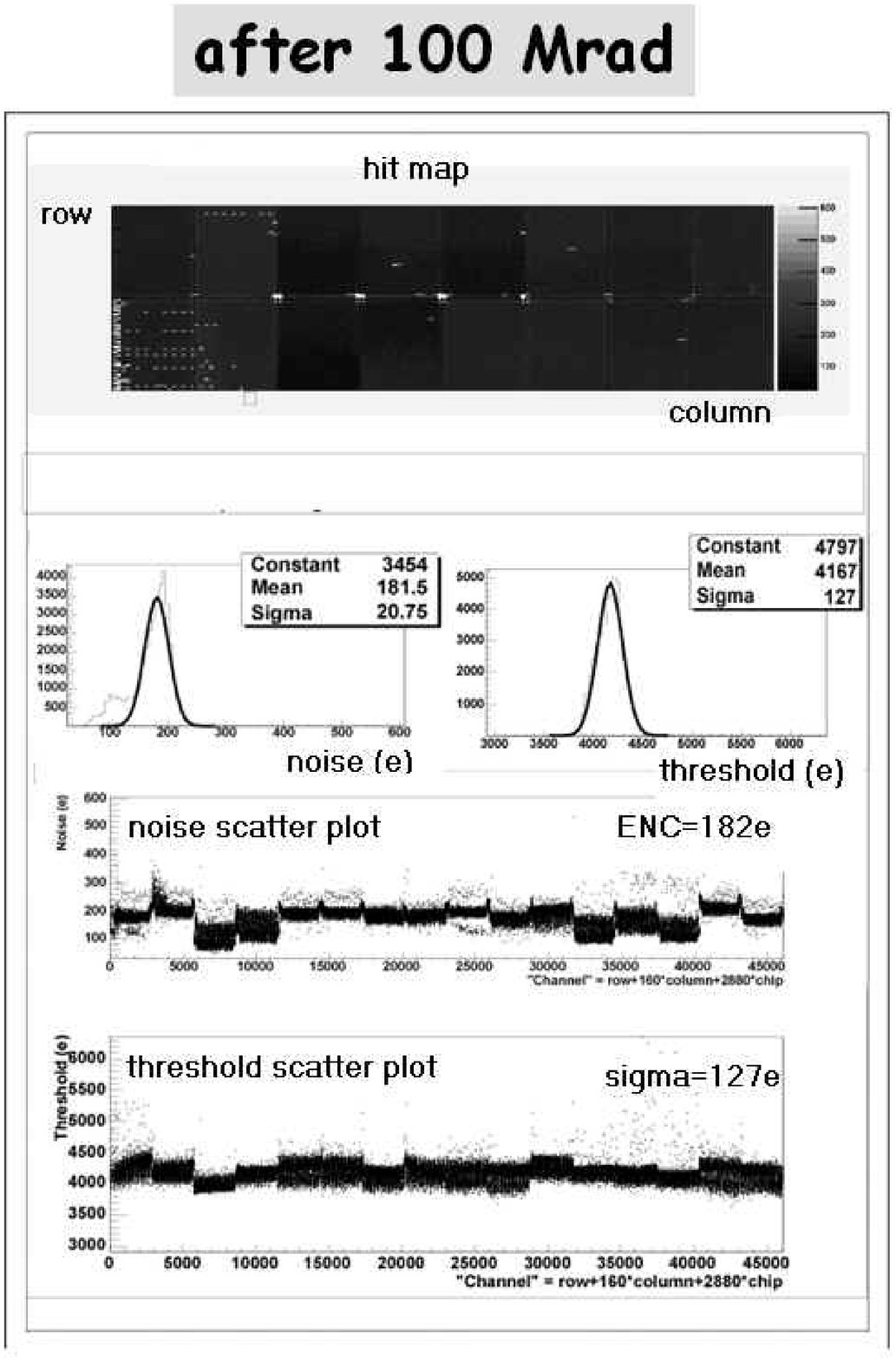}
\vskip 0.5cm
\includegraphics[width=0.38\textwidth]{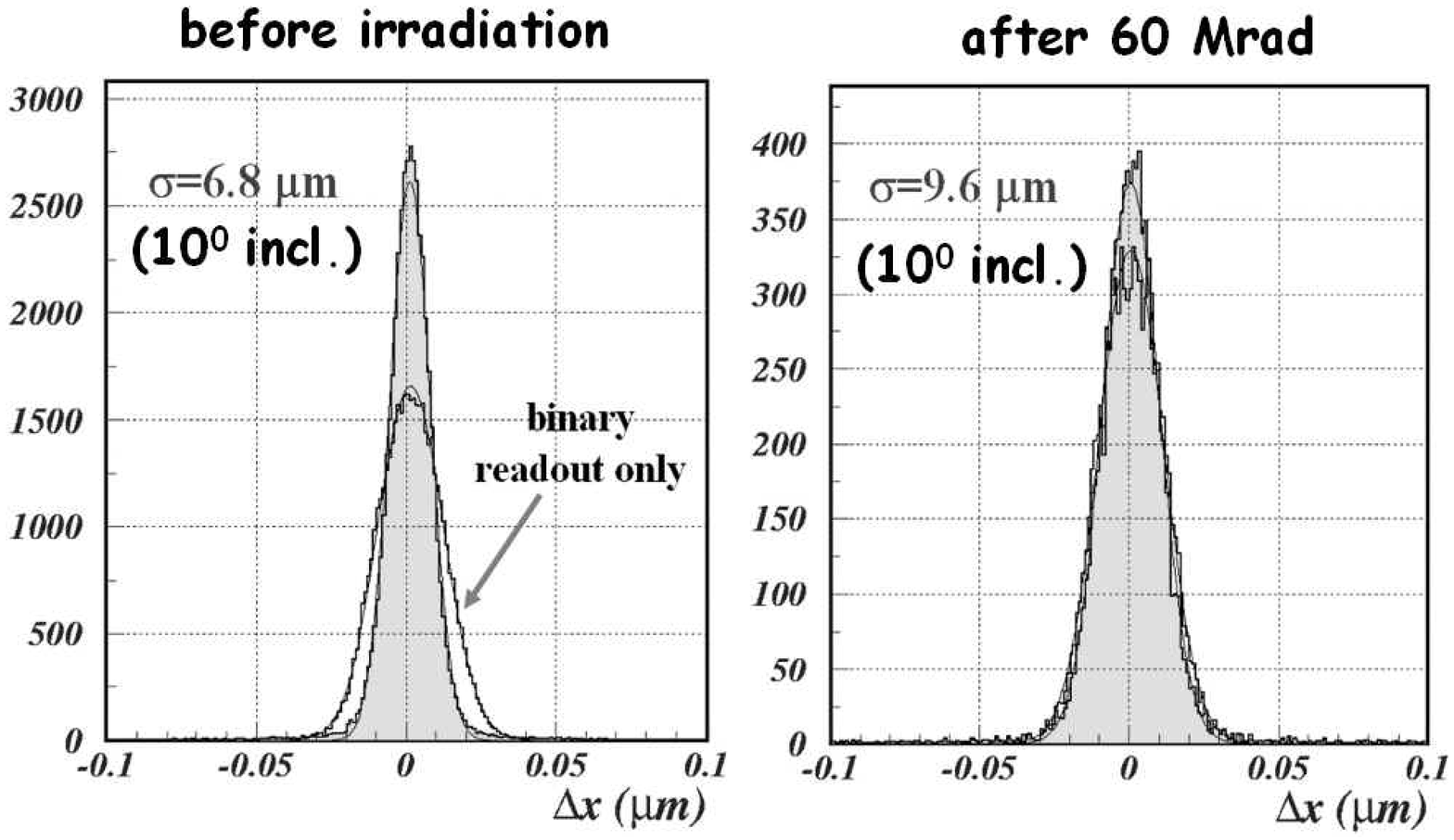}
\vskip 0.3cm
\includegraphics[width=0.38\textwidth]{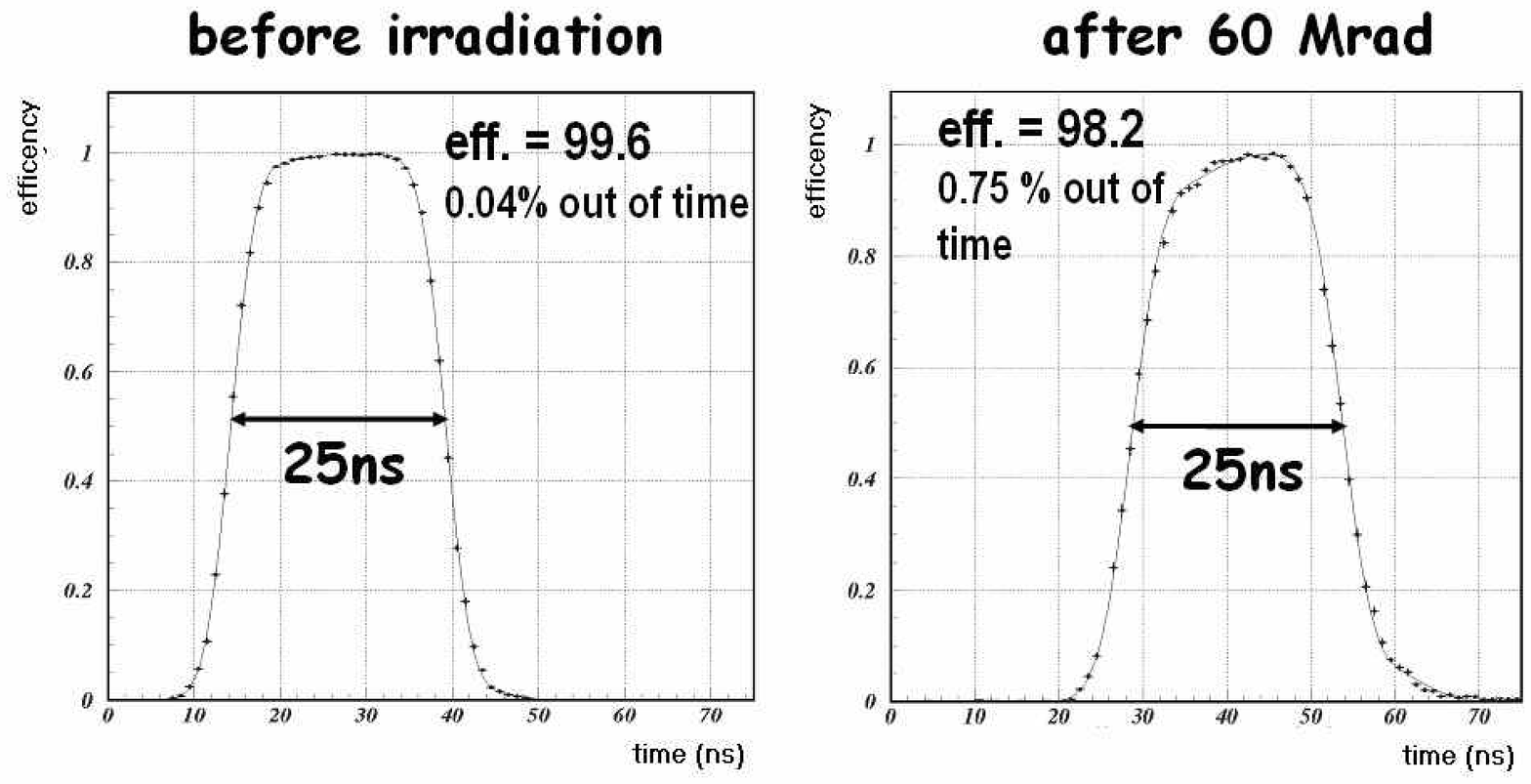}
\end{center}
  \caption{Comparisons of ATLAS pixel modules before and after
  irradiation to doses up to 100 Mrad.
  Hit map, noise and threshold dispersions (top), spatial
  resolution in the 50 $\mu$m direction of the pixels at 10$^o$
  incidence angle (center), and
  the hit efficiency (bottom). The in-time efficiency of a hit to be earlier than 25 ns
  is determined
  in test beams relative to a fixed delay of the trigger counters.
  The highest point of the plateau shows the in-time efficiency. The width
  of the plateau characterizes the available margin during operation.}
  \label{irrad}
\end{figure}
\begin{figure}[b]
\begin{center}
\includegraphics[width=0.45\textwidth]{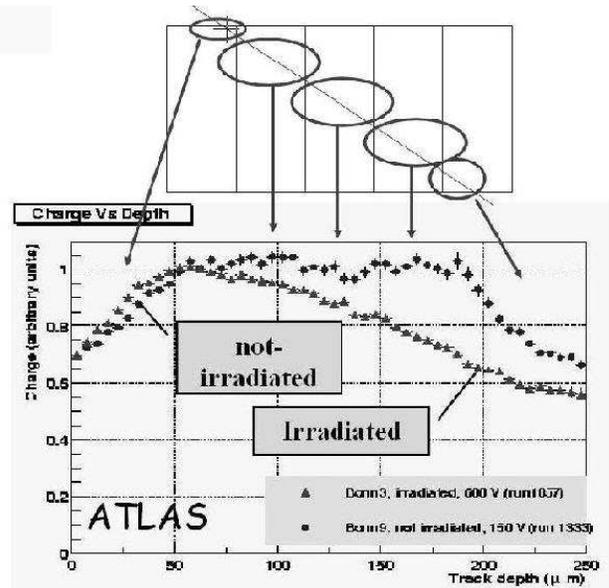}
\end{center}
  \caption{Charge collection efficiency as a function of the depth
  of the track traversing the detector.}
  \label{trapping}
\end{figure}

Most challenging at LHC is the requirement on radiation tolerance
of the pixel detector which is exposed to (mostly) pion fluences
of 10$^15$~n$_eq$/cm$^2$ or 500~kGy during 10 years of LHC
operation. The advancement of oxygenated silicon and deep
submicron chip technology made a long life time in such an
environment possible. Figures \ref{irrad}(a)-(c) show the
comparison of critical performance figures before and after
irradiation of ATLAS pixel modules. In parts the received dose of
the modules was well in excess of that expected for 10 years
operation at the LHC. After irradiation to 500~kGy the mean
collected charge fraction has been measured to be $\sim$80$\%$ and
the in-time efficiency, i.e. the efficiency for hit detection
within 20ns after the bunch crossing, is 97.8$\pm$0.1$\%$. An
important characterization figure is the in-time efficiency, which
in test beams, at which the arrival of the beam particles is
asynchronous to the system clock, can be determined by plotting
the hit efficiency as a function of the delay between the arrival
time and the clock edge, measured by a TDC. It is mandatory that a
high efficiency is reached somewhere inside a plateau. The width
and flatness of the plateau is a measure of some kind of margin
that exists. The plateau width decreases after irradiation from
originally 14~ns to 9.7~ns. Figure~\ref{trapping} shows a
comparison between irradiated and not irradiated pixel sensors
regarding the amount of trapping. By measuring the charge
collection efficiency under inclined angles different depths for
the charge deposition in the sensor can be addressed and hence
trapping can be studied as shown by Fig.~\ref{trapping}. After 10
years at the LHC the charge collection efficiency is about 80$\%$.
The charge yield as a function of depth can be translated into an
electron carrier life time of $\tau$ = 4.1$\pm$ 0.6 ns.

\subsection{Support structures and total thickness}
For ALICE the demands imposed by the physics are different than
for ATLAS and CMS. While the track density ($\sim$80 hits/cm$^2$)
with 8000 charged particles per rapidity interval for central
heavy ion collisions is truly formidable, the radiation level of
5~kGy or 6$\times$10$^{12}$n$_{eq}$/cm$^2$, due to the much lower
collision rate, is much lower than for pp collisions. As a
consequence cooling to temperatures below 0$^\circ$C as for CMS
and ATLAS is not mandatory, but instead very thin materials for a
small total radiation length are aimed for. The reduced cooling
requirement (24$^\circ$ C) allows the use of a very light weight
structure with 40$\mu$m wall thickness PHYNOX tubes and a total
contribution of 0.3$\%$X$_0$. Together with thin sensors (200
$\mu$m) and chips (150 $\mu$m) an ALICE module (1.28 cm $\times$
7.0 cm, 5 readout chips bonded to one sensor) arrives at a total
radiation length per layer of only 0.9$\%$.

\begin{figure}[htb]
\begin{center}
\includegraphics[width=0.5\textwidth]{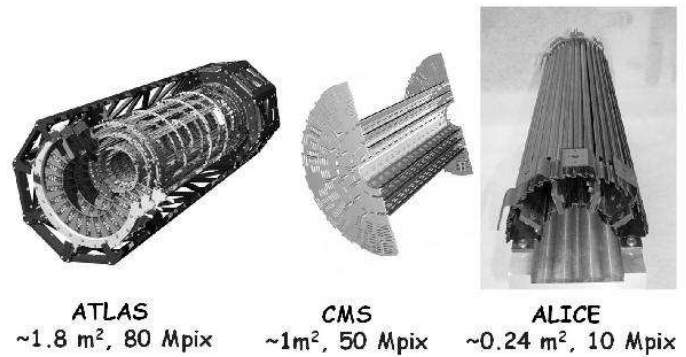}
\end{center}
  \caption{Support structures of the ATLAS (carbon-carbon), CMS (carbon-fibre),
  and ALICE (carbon-fibre) pixel detector global support structures. .}
  \label{global_supports}
\end{figure}

\subsection{LHC pixels put to test in NA60}
\begin{figure}[htb]
\begin{center}
\includegraphics[width=0.5\textwidth]{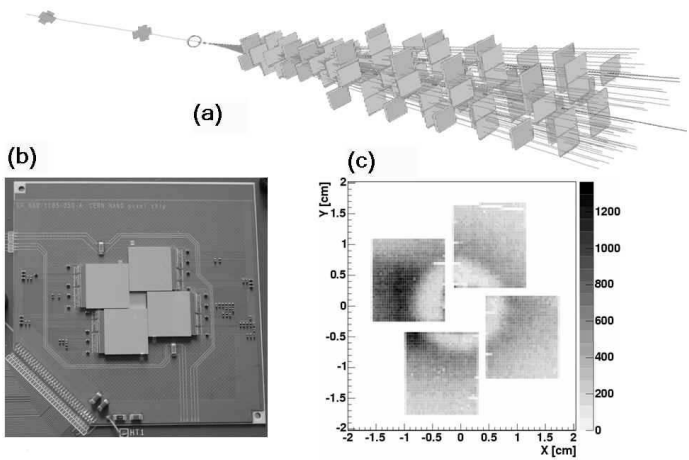}
\end{center}
  \caption{Pixel detector tracker in the NA60 experiment (top) consisting in total of
    16 track hit planes, four of these using ATLAS production modules. The pixel
    planes(bottom left)
    were operated through partial type-inversion, with the results
    demonstrated in the hit-multiplicity plot (bottom right).
  }
  \label{NA60}
\end{figure}

The CERN heavy ion experiment NA60~\cite{NA60_Keil} has used
LHC-type pixel detectors for the first time in a running
experiment. The setup of the NA60 pixel tracker is shown in
Fig.~\ref{NA60} (top). For the initial running the ALICE-LHCb chip
was used. Eight 4-chip (Fig.~\ref{NA60}(bottom left)) and eight
8-chip planes provide track reconstruction with 12 pixel hits on a
track. The sensors have been exposed to a radiation dose of 120
kGy and were operated through type inversion. Due to the
inhomogeneous irradiation the inner part of the planes has
received a larger dose than the outer, which is demonstrated by
the hit multiplicity pattern taken with a lowered bias voltage in
Fig. \ref{NA60}(bottom right). The improvement in physics from the
operation of the pixel vertex detector has been reported in
\cite{NA60_Keil}. The Indium target position can be resolved to
20$\mu$m in the direction transverse to the beam and 200$\mu$m in
the longitudinal direction. The meson resonances $\rho$ and
$\omega$ could be detected with a resolution of 23 MeV in the
di-muon invariant mass. In 2005, data for p-nucleus running at a
beam intensity of 2$\times$10$^9$ p/burst and an interaction rate
of one per 25 ns has been taken. In order to cope with such
LHC-like data rates, four planes using ATLAS production modules
were added to the setup.

\section{Spin-off from hybrid pixel detectors into other fields}
\subsection{X-ray imaging using counting pixel detectors}
Spin-off from hybrid pixel detectors in particle physics has most
directly arisen in imaging applications as detectors that
accumulate the incident radiation by the counting of individual
radiation quanta in every pixel cell. This technique offers many
features which are very attractive for X-ray imaging: full
linearity in the response function, in principle an infinite
dynamic range, optimal exposure times and a good image contrast
compared to conventional film-foil based radiography, thus
avoiding over- and under-exposed images. The analog part of the
pixel electronics is in parts close to identical to the one for
LHC pixel detectors while the periphery has been replaced by
counting circuitry \cite{PeFicounter}. The same principle is also
used for protein-crystallography with synchrotron radiation
\cite{Graafsma_Portland,3Dwestbrook}.

The challenges which are to be addressed in order to be
competitive with integrating systems are: high speed ($>$ 1 MHz)
counting with a range of at least 15 bits, operation with very
little dead time, low noise and particularly low threshold
operation with small threshold dispersion values. In particular,
the last item is important in order to allow homogeneous imaging
of soft X-rays of energies in the energy range below 10 keV. It is
also mandatory for a differential energy measurement, realized so
far as a double threshold with energy windowing logic
\cite{MPEC-windowing1,MPEC-windowing2,MEDIPIX2}, which can enhance
the contrast of an image as the shape of the X-ray energy spectrum
is different behind different absorbers (e.g. bone or soft
tissue). Finally, for radiography, high photon absorption
efficiency is mandatory, requiring the use and development of
high-Z sensors and their hybridization.

\begin{figure}
\begin{center}
\includegraphics[width=0.24\textwidth]{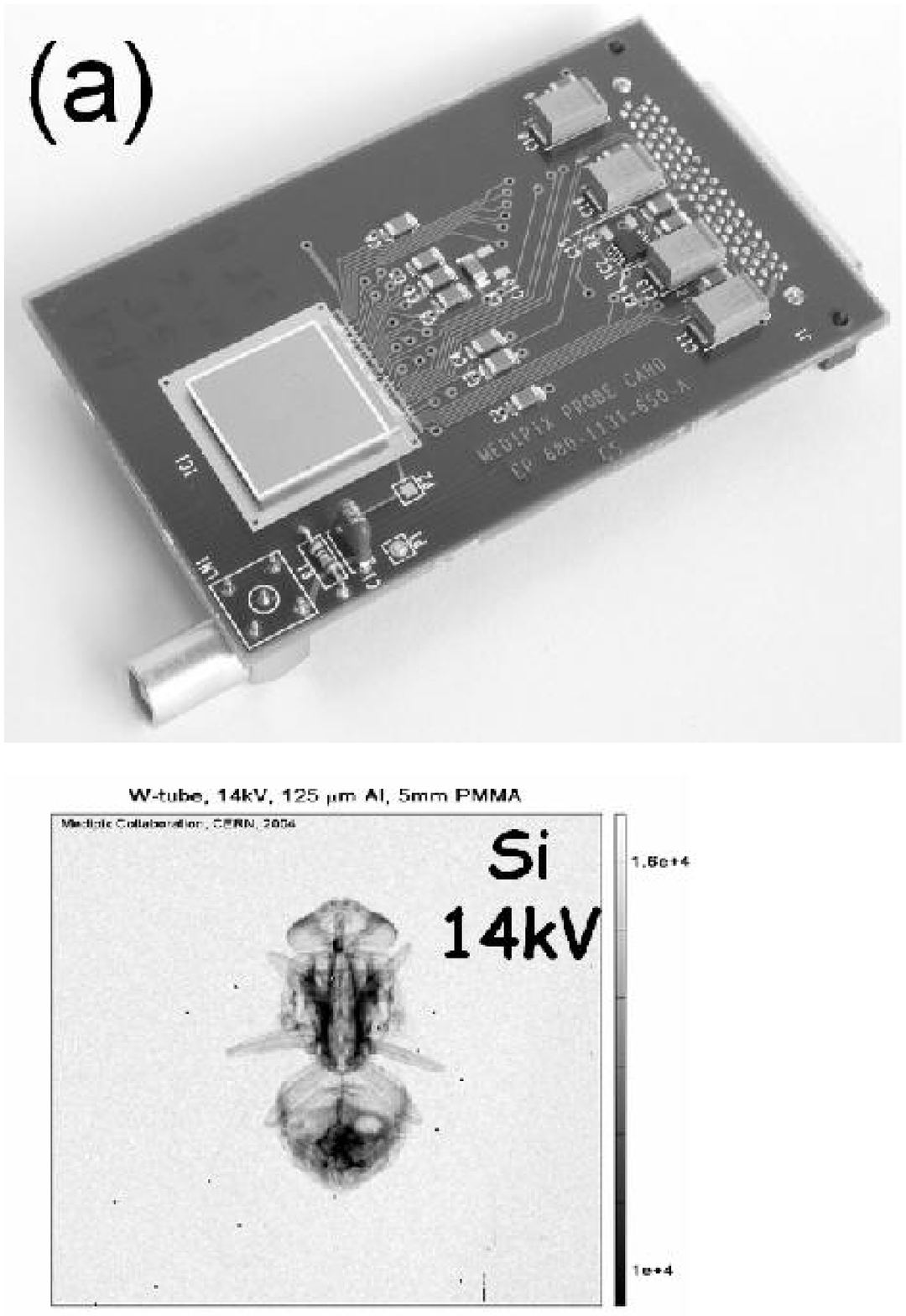}
\includegraphics[width=0.24\textwidth]{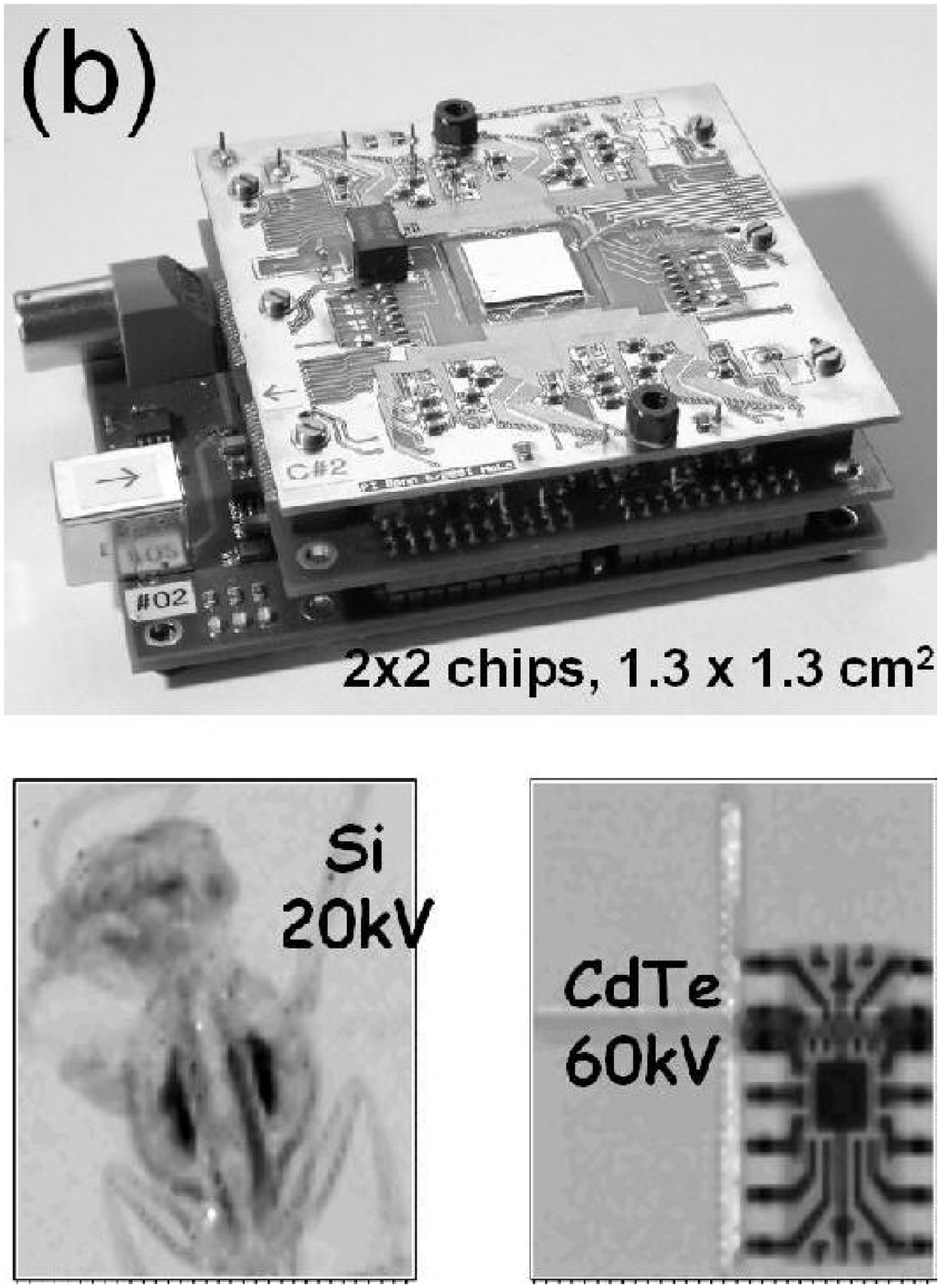}
\end{center}
\caption[]{(a) MEDIPIX2 counting pixel chip module (14x14 mm$^2$,
55$\times$55$\mu$m pixel size) with Si sensor~\cite{MEDIPIX2}
(top) and an image of a bee (bottom); (b) MPEC 2x2 multi chip
module with a CdTe sensor~\cite{MPEC_Portland} (top) and images of
a hornet and a transistor housing (bottom).} \label{counting}
\end{figure}

The MEDIPIX collaboration~\cite{MEDIPIX2} uses the MEDI\-PIX2 chip
with $256 \times 256$, 55$\times $55 $\mu$m$^2$ pixels fabricated
in a $0.25 \mu$m technology, energy windowing via two tunable
discriminator thresholds, and a 13 bit counter. The maximum count
rate per pixel is about 1 MHz. Fig. \ref{counting}(a) a single
chip module together with an image of a bee taken using 14 keV
X-rays~\cite{MEDIPIX2}. A Multi-Chip module with 2x2 chips using
high-Z CdTe sensors is shown in Fig.~\ref{counting}(b) using the
MPEC chip \cite{MPEC_Portland}, together with two X-ray images.
The MPEC chip features $32 \times 32$ pixels
(200$\times$200$\mu$m$^2$), double threshold operation, 18-bit
counting at $\sim$1 MHz per pixel as well as low noise values
($\sim$120e with CdTe sensor) and threshold dispersion ($21$e
after tuning) \cite{MPEC_ref,MPEC_Portland}. A technical issue
here is the bumping of individual die CdTe sensors which has been
solved using Au-stud bumping with Indium-filling \cite{MPEC_CdTe}.

\subsection{Counting pixels in protein crystallography}
\begin{figure}
\begin{center}
\includegraphics[width=0.22\textwidth]{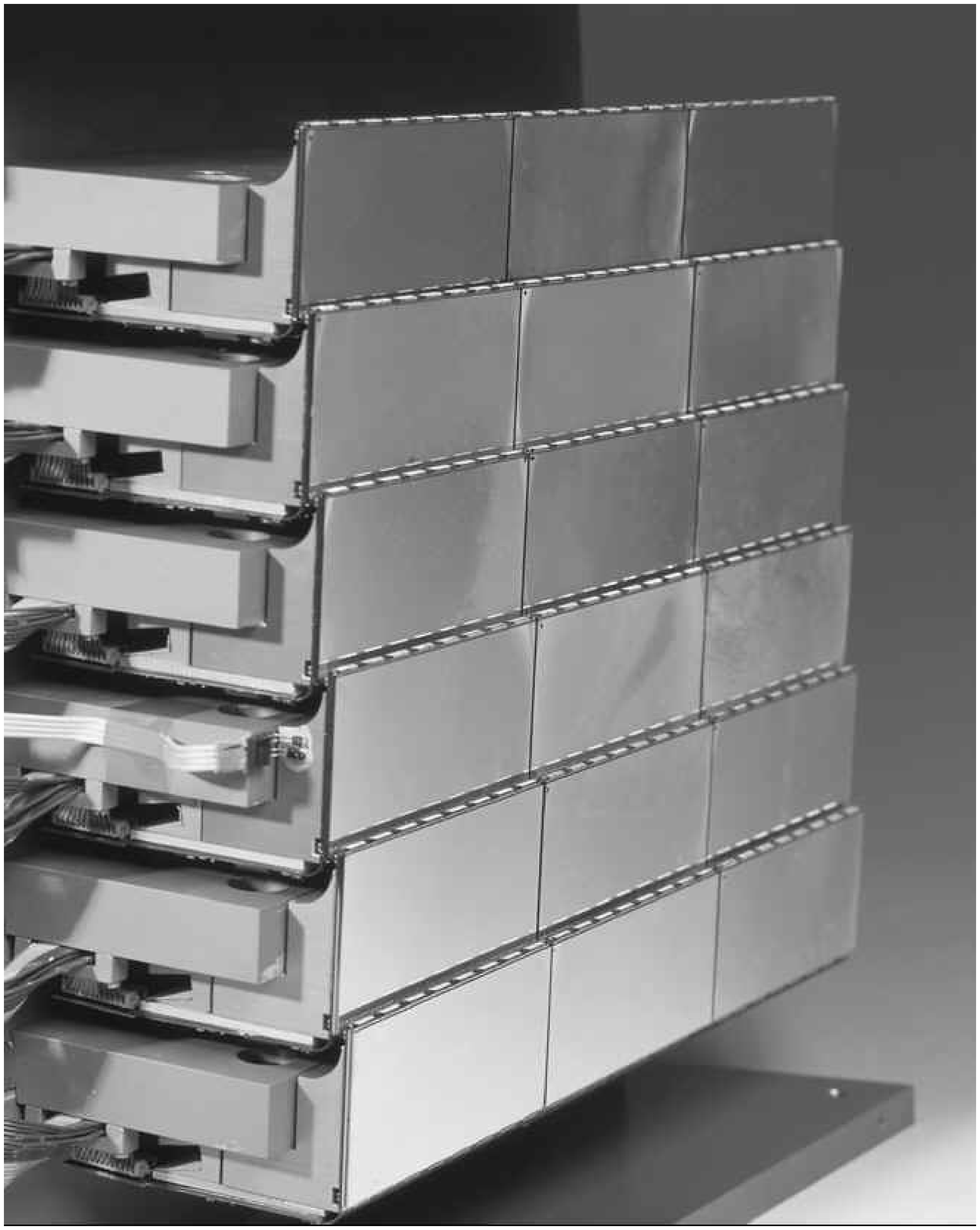}
\includegraphics[width=0.22\textwidth]{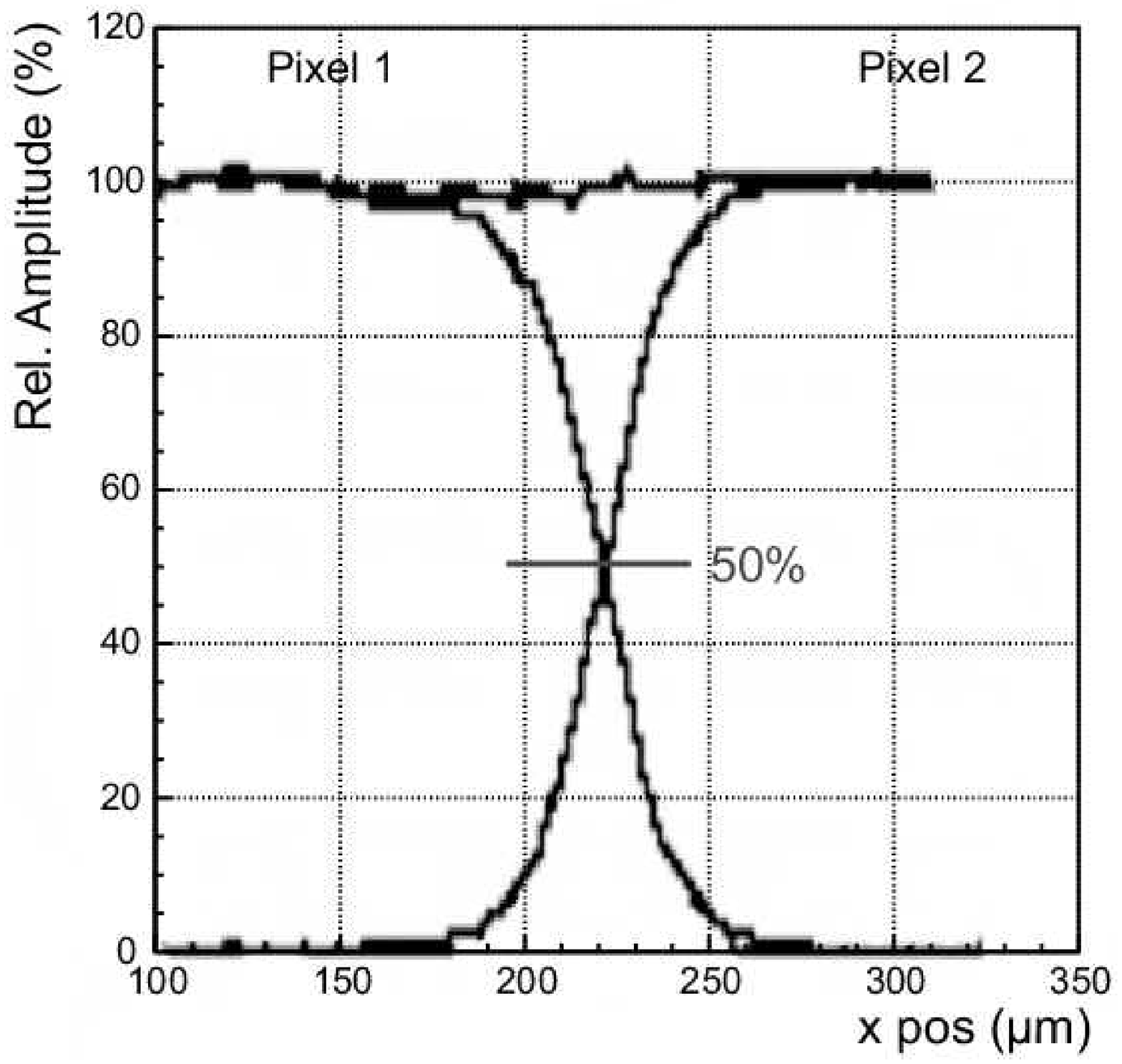}
\vskip 0.5cm
\includegraphics[width=0.30\textwidth]{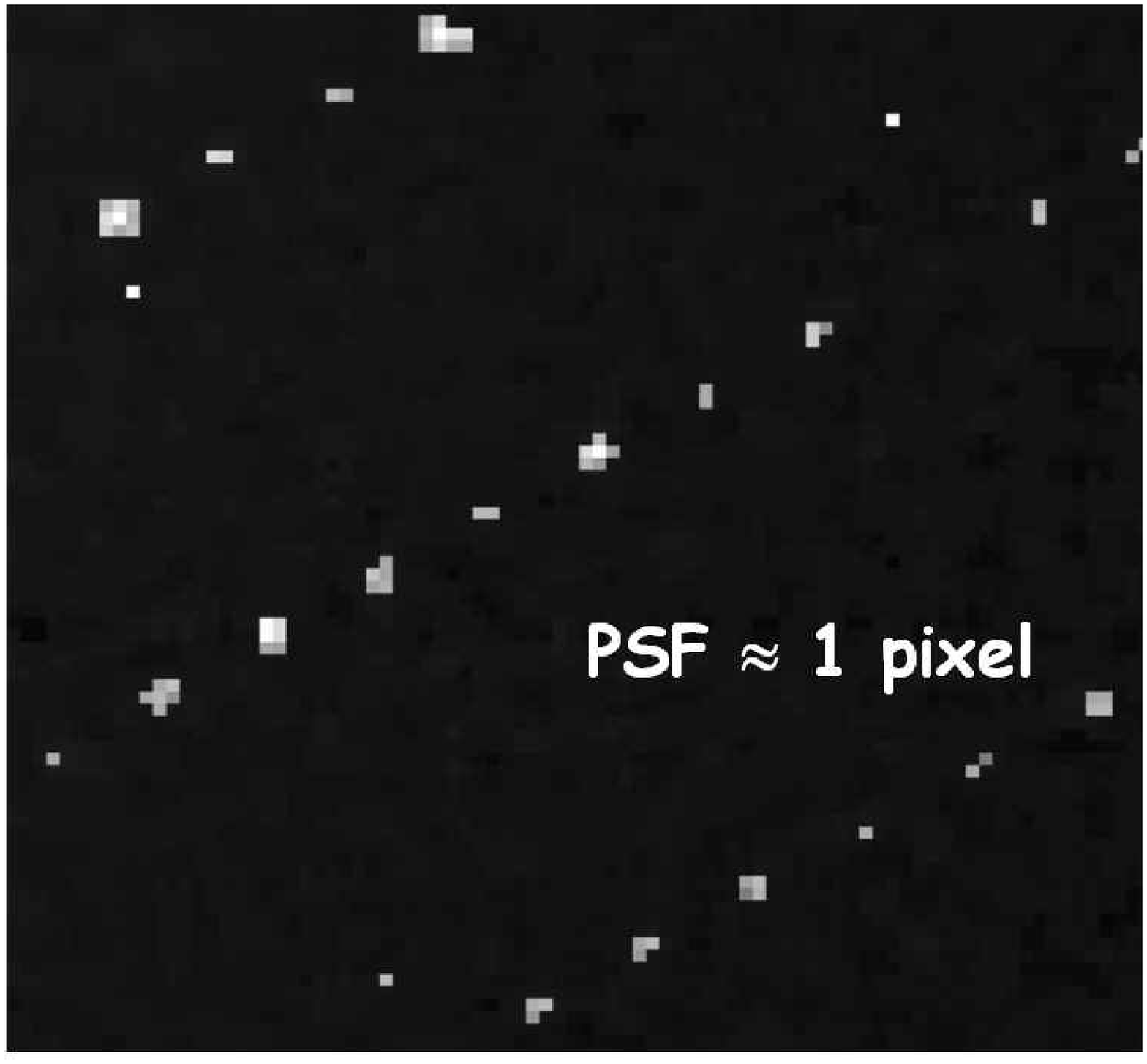}
\includegraphics[width=0.18\textwidth]{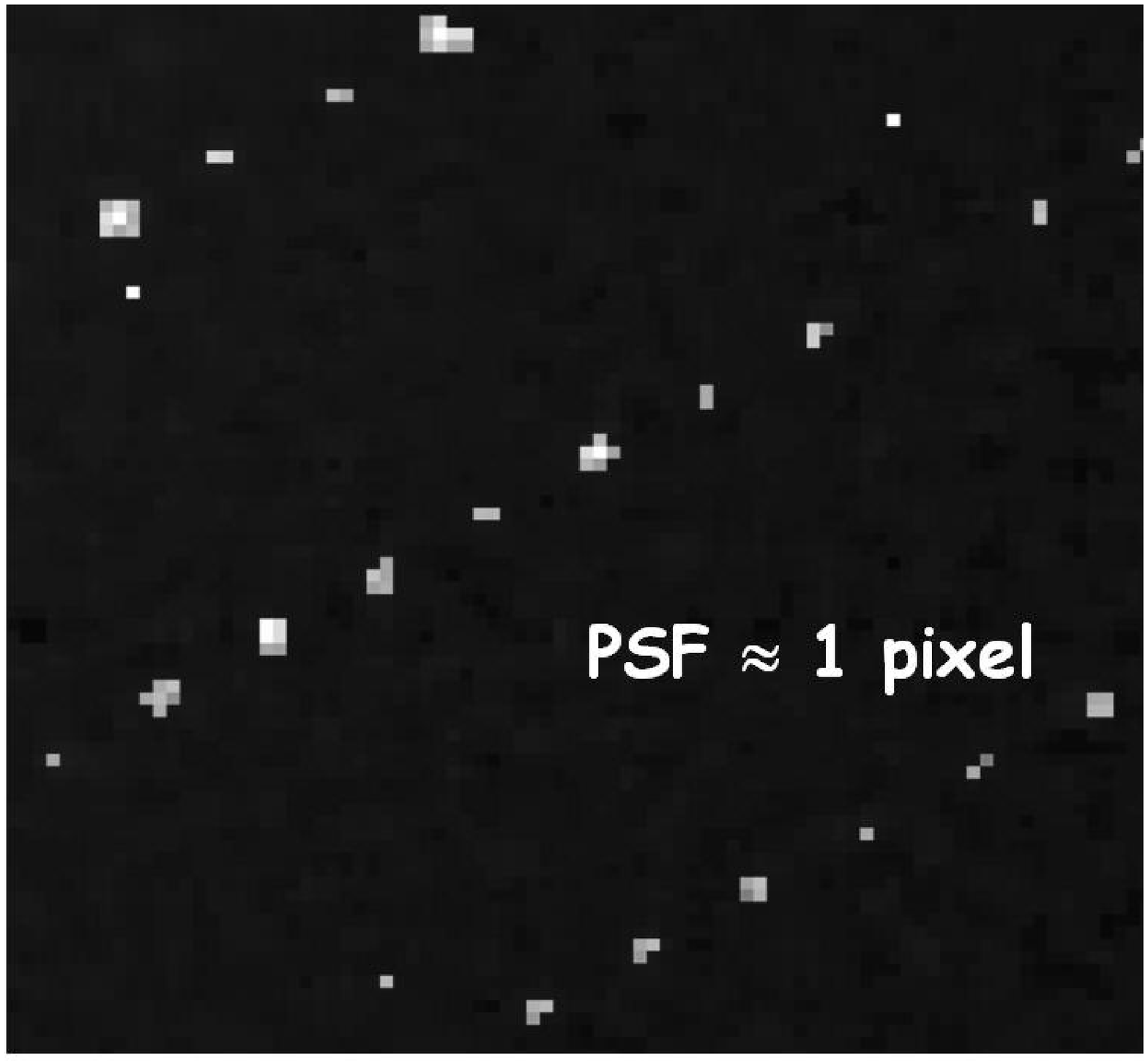}
\includegraphics[width=0.25\textwidth]{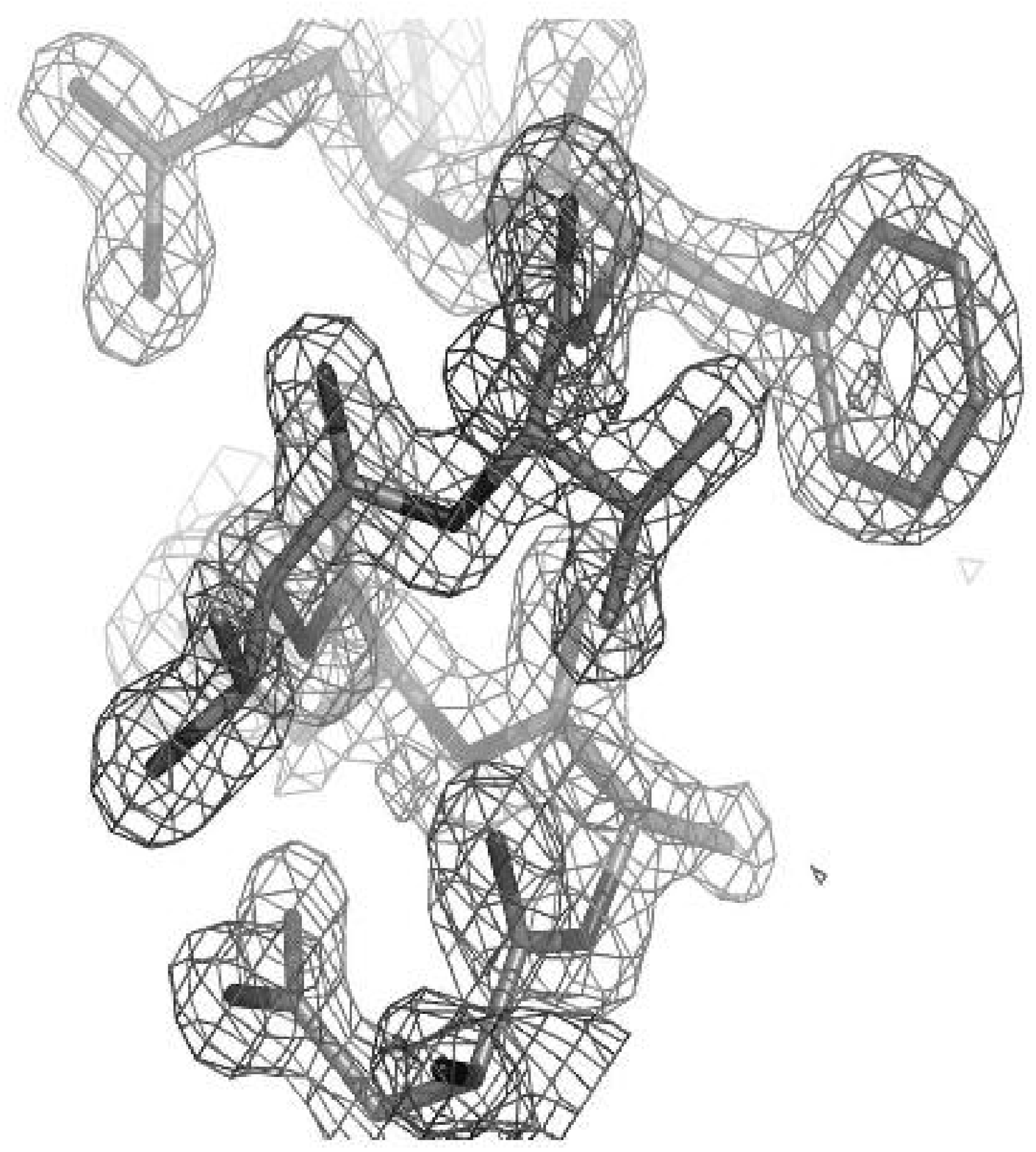}
\end{center}
\caption[]{(top left) Photograph of the 20x24 cm$^2$ large PILATUS
1M detector (PSI) for protein crystallography using counting
hybrid pixel detector modules; (top right) delicate threshold
tuning at the borders in between pixels; (center) flat field image
of a module; (bottom left) Bragg spots of an image taken with
PILATUS 1M \cite{PILATUS_Portland} are often contained in one
pixels; (bottom right) reconstructed electron density map of
thaumatin molecule} \label{crystallography}
\end{figure}

In {\it protein crystallography} with synchrotron
radiation~\cite{Graafsma_Portland} the challenge is to image many
thousands of Bragg spots from X-ray photons with energies of
$\sim$12 keV (corresponding to resolutions at the 1$\AA$ range) or
higher, scattering off protein crystals. This must be accomplished
at a high rate ($\sim$1-1.5 MHz/pixel) and by systems with a high
dynamic range. The typical spot size of a diffraction maximum is
$100-200 \mu$m, calling for pixel sizes in the order of $100-300
\mu$m. The high linearity of the hit counting method and the
absence of so-called "blooming effects", i.e. the response of
non-hit pixels in the close neighborhood of a Bragg spot, makes
counting pixel detectors very appealing for protein
crystallography experiments. Counting pixel developments are made
for the ESRF (Grenoble, France)~\cite{XPAD2} and the SLS (Swiss
Light Source at the Paul-Scherrer Institute, Switzerland) beam
lines. A photograph of the PILATUS 1M detector~\cite{PILATUS} at
the SLS ($\sim$ 10$^6$ $217 \mu$m$\times 217 \mu$m pixels, 18
modules, 20$\times$24 cm$^2$ area) is shown in
Fig.~\ref{crystallography}(a). A systematic limitation and
difficulty is the problem that homogeneous hit/count responses in
all pixels, also for hits at the pixel boundaries or between
pixels where charge sharing plays a role must be maintained by
delicate threshold tuning (cf~Fig.~\ref{crystallography}(b)).
Fig.~\ref{crystallography}(c) shows a flat-field image obtained
with a PILATUS module, which demonstrates that this principle
problem can be overcome~\cite{henrich_Pix2005}.
Figure~\ref{crystallography}(d), shows some Bragg spots obtained
with a 10s exposure to 12 keV synchrotron
X-rays~\cite{PILATUS_Portland}. Some spots are contained in only
one pixel, others spread over a few pixels due to charge sharing.
This demonstrates the intrinsically good point resolution of the
system. Figure~\ref{crystallography}(e) is a reconstructed
electron density map of the thaumtin
crystal~\cite{henrich_Pix2005}. Alternative developments which aim
to improve the active/inactive area ratio for
protein-crystallography X-ray detection are so-called 3-D silicon
sensors (strip or pixels)~\cite{3D-parker}. A detailed account can
be found in~\cite{parker_PIX2005}.

\section{Challenges imposed by a Super-LHC}
The radiation levels expected at an LHC upgrade, called
Super-LHC or SLHC, are a factor of ten higher than at the LHC,
i.e. to 10$^{16}$n$_{eq}$/cm$^2$. There are mainly three effects
as a consequence~\cite{moll_PIX2005}.
\begin{enumerate}
\item[1.] A change of the effective doping concentration (higher
depletion voltage necessary, under-depletion) \item[2.] An
increase of leakage current (increase of shot noise, thermal
runaway) \item[3.] An increase of charge carrier trapping (loss of
charge)
\end{enumerate}
Several routes to cope with this are being pursued, among them the
development of even more radiation hard silicon based on
oxygenated float-zone (DOFZ), Czochralski (Cz) and epitaxial
silicon~\cite{moll_PIX2005}. For this review I would like to
address in this context two new approaches which are more linked
to pixel detectors: diamond pixel detectors and 3D-silicon
devices.

\subsection{Diamond Pixels}
\begin{figure}[h]
\begin{center}
\includegraphics[width=0.5\textwidth]{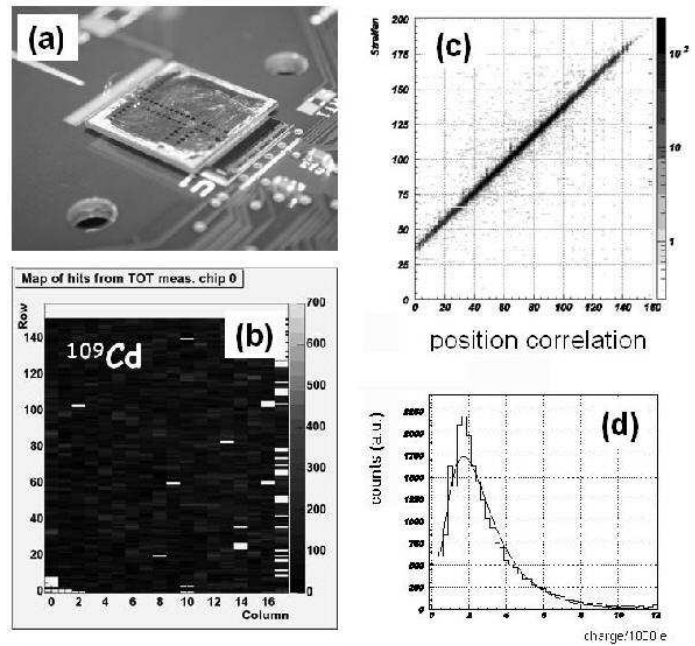}
\end{center}
\caption[]{(a) Single chip diamond pixel module using ATLAS
front-end electronics, (b) hit map obtained by exposure to a
$^{109}$Cd radioactive source (22 keV $\gamma$), (c) scatter plot
of position correlation between the diamond pixel detector and a
reference beam telescope, and (d) measured Landau distribution in
a CVD-diamond pixel detector.} \label{diamond-single}
\end{figure}
\begin{figure}
\begin{center}
\includegraphics[width=0.5\textwidth]{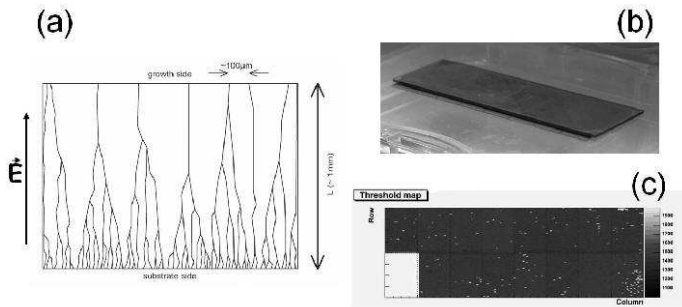}
\end{center}
\caption[]{(a) Grain structure of CVD-diamond sensors. (b) a full
size CVD-diamond module from a CVD diamond wafer bump bonded to 16
ATLAS FE-chips (on the bottom), (c) threshold map after tuning of
the module showing its full functionality.}\label{diamond-grains}
\end{figure}

CVD-Diamond as a sensor material has been developed by the CERN
R$\&$D group RD42 for many years~\cite{RD42}. Charge collection
distances approaching 300 $\mu$m has also triggered the
development of a hybrid pixel detector using diamond as sensors
\cite{diamond-pixels,kagan_PIX2005}. The non-uniform field
distribution inside CVD-diamond, which originates from the grain
structure in the charge collecting bulk
(cf.~Fig.~\ref{diamond-grains}(a)) introduces polarization fields
inside the sensor due to charge trapping at the grain boundaries
which superimpose on the biasing electric field. This results in
position dependent systematic shifts in the track reconstruction
with a typical average grain size of 100$\mu$m -
150$\mu$m~\cite{lari04}. Diamond sensors with charge collection
distances in excess of 300$\mu$m have been fabricated and
tested~\cite{Kagan_Hiroshima}. Single chip pixel modules as well
as a full size wafer scale 16-chip module assembled using ATLAS
front-end chips have been built and tested.
Figure~\ref{diamond-single}(a) and (b) show the diamond pixel
detector and a hit response pattern obtained by exposing the
detector to a $^{109}$Cd source of 22 keV $\gamma$ rays, which
deposits approximately 1/4 of the charge of a minimum ionizing
particle. The single chip module has been tested in a high energy
(180 GeV) pion beam at CERN, the module in a $\sim$4 GeV electron
beam at DESY. Figure~\ref{diamond-single}(c) and (d) show position
correlation and the charge distribution of the diamond pixel
detector in a high energy beam, respectively. A spatial resolution
of $\sigma$ = 12$\mu$m has been measured with the single chip
module at high energies with 50$\mu$m pixel pitch. A technical
challenge to produce a wafer scale module lies in the
hybridization process, i.e. bump deposition and flip-chipping.
Figure~\ref{diamond-grains} shows the 16-chip diamond module
(Fig.~\ref{diamond-grains}(b)) and its tuned threshold map
(Fig.~\ref{diamond-grains}(c)) with a very small dispersion of
only 25e$^-$ and good bump yield homogeneity. One chip was damaged
during test beam by electrostatic discharge. The rms of the
position residuals was measured to 24~$\mu$m in the DESY 6 GeV
beam~\cite{kagan_PIX2005}. This value is dominated by the multiple
scattering contribution from the beam telescope.
%

\subsection{3D silicon sensors}
So-called 3D silicon detectors have been
developed~\cite{3D-parker} to overcome several limitations of
conventional planar Si-pixel detectors, in particular in high
radiation environments, in applications with inhomogeneous
irradiation and in applications which require a large
active/inactive area ratio such as protein
crystallography~\cite{3Dwestbrook}.
\begin{figure}
\begin{center}
\includegraphics[width=0.35\textwidth]{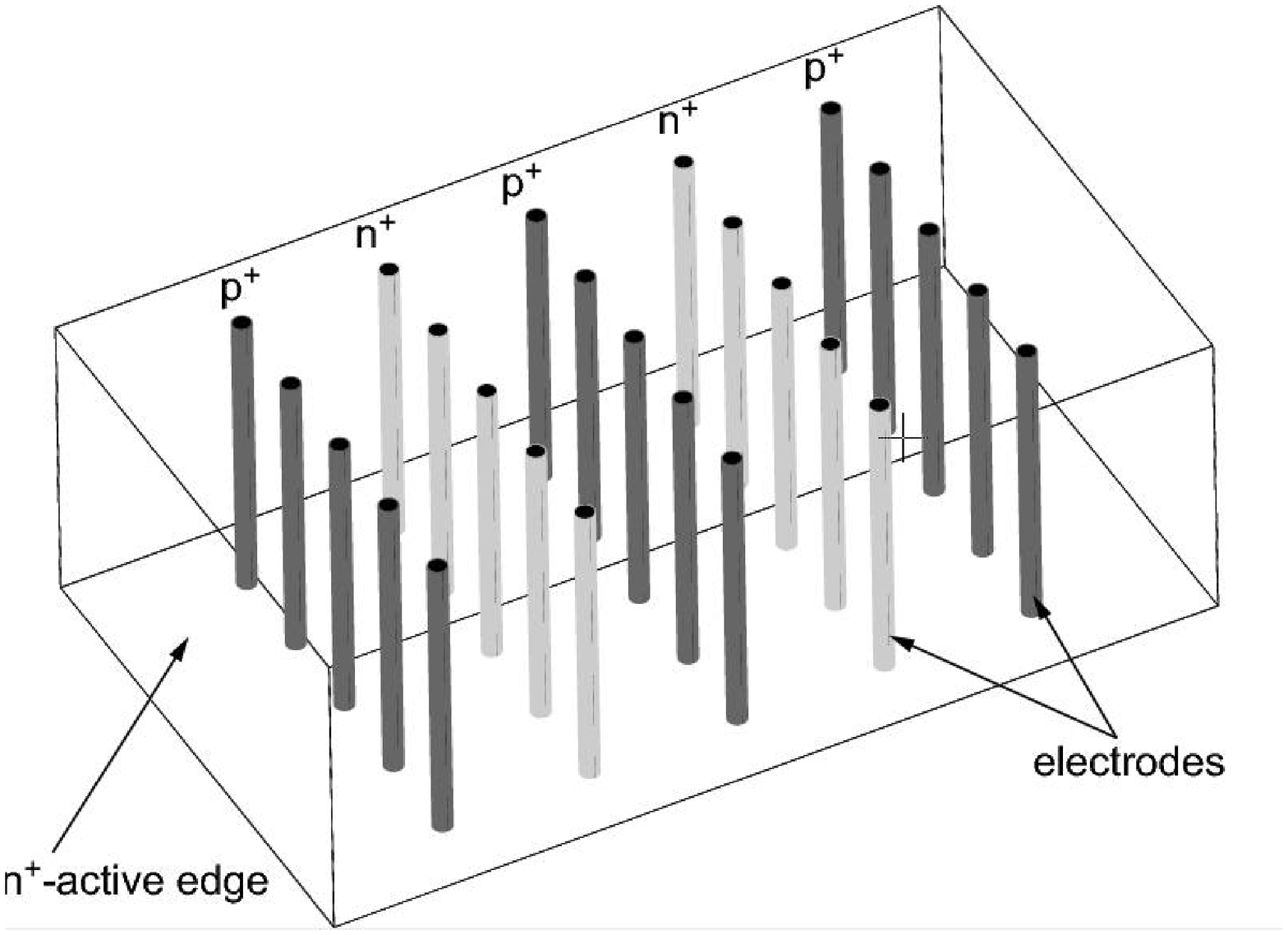}
\vspace{0.4cm}
\includegraphics[width=0.35\textwidth]{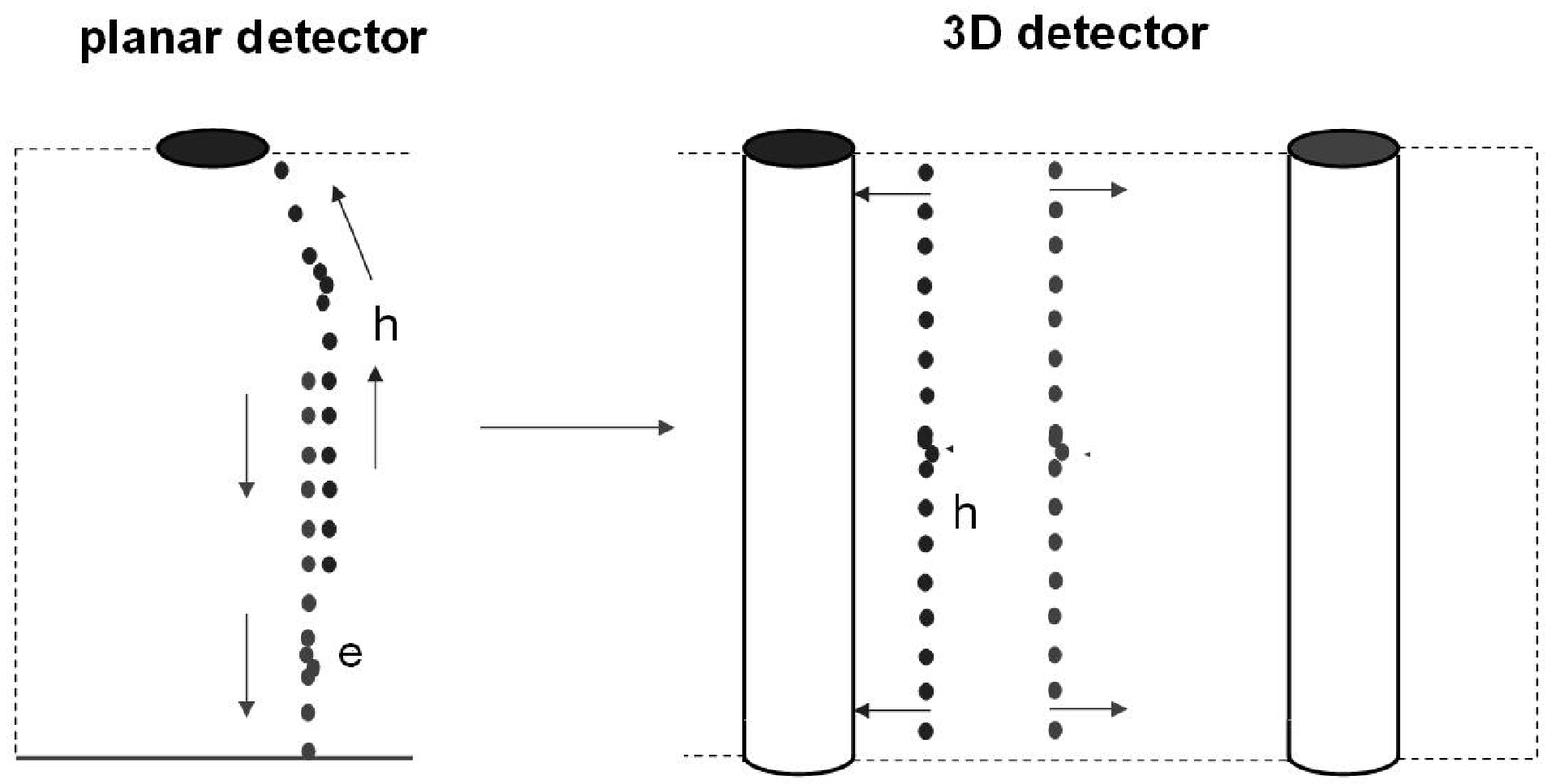}
\end{center}
\caption[]{(top) Schematic view of a 3D silicon detector, (bottom
left) comparison of the charge collection in a conventional planar
electrode silicon detector, (bottom right) a 3D-silicon detector.}
\label{3D}
\end{figure}
A a 3D-Si-structure (Fig.~\ref{3D}(a)) is obtained by processing
the n$^+$ and p$^+$ electrodes into the detector bulk rather than
by conventional implantation on the surface. This is done by
combining VLSI and MEMS (Micro Electro Mechanical Systems)
technologies. Charge carriers drift inside the bulk parallel to
the detector surface over a short drift distance of typically
50$\mu$m. Another feature is the fact that the edge of the sensor
can be a collection electrode itself thus extending the active
area of the sensor to within few $\mu$m to the edge. Edge
electrodes also avoid inhomogeneous fields and surface leakage
currents which usually occur due to chips and cracks at the sensor
edges. The main advantages of 3D-silicon detectors, however come
from a different way of charge collection and the fact that the
electrode distance is short (50$\mu$m) in comparison to
conventional planar devices at the same total charge, This results
in a fast (1-2 ns) collection time, low ($<$ 10V) depletion
voltage and, with edge electrodes in addition, a large
active/inactive area ratio of the device (cf. Fig.~\ref{3D}(b)).

The technical fabrication is much more involved than for planar
processes and requires a bonded support wafer and reactive ion
etching of the electrodes into the bulk. A compromise between 3D
and planar detectors, so called planar-3D detectors maintaining
the large active area, use planar technology but with edge
electrodes~\cite{3Dplanar}, obtained by diffusing the dopant from
the deeply etched edge and then filling it with poly-silicon.
Prototype detectors using strip or pixel electronics have been
fabricated and show encouraging results with respect to speed (3.5
ns rise time) and radiation hardness ($\gg$ 10$^{15}$
protons/cm$^2$)~\cite{3DPortland}. 3D-pixel detectors with
LHC-type frontend electronics have not been successfully built
yet, although the hybridization imposed no problem compared to
standard hybrid pixel devices. A fabrication of 3D-pixel
structures adapted to the ATLAS FE-chip is
underway~\cite{parker2005}.

\section{Monolithic and Semi-Monoli\-thic Pixel Detectors}
Monolithic pixel detectors, in which amplifying and
logic circuitry as well as the radiation detecting sensor are one
entity, are in the focus of present developments. To reach this
ambitious goal, optimally using a commercially available and cost
effective technology, would be another breakthrough in the field.
The present developments have been much influenced by R$\&$D for
vertex tracking detectors at future colliders such as the
International Linear $e^+e^-$ Collider (ILC) \cite{TESLA-TDR}.
Very low ($\ll$1$\%$ X$_0$) material per detector layer, small
pixel sizes ($\sim$20$\mu$m$\times 20 \mu$m) and a high rate
capability (80 hits/mm$^2$/ms) are required, due to the very
intense beamstrahlung of narrowly focussed electron beams close to
the interaction region, which produce electron positron pairs in
vast numbers. High readout speeds with typical line rates of
several $10$ MHz and a 40$\mu$s frame readout time are necessary.

At present, two developments have already reached some level of
maturity: so called CMOS active pixels and DEPFET pixels. Other
promising approaches, not mentioned in this review, are amorphous
a-Si:H layers for charge collection superimposed on standard CMOS
ASICS~\cite{jarron02} as well as so called
SOI-sensors~\cite{SOI_2005}, which use a high-ohmic Si-substracte
with full charge collection wafer-bonded to a CMOS electronics
layer. Both active layers (sensor and CMOS) are isolated by an
insulating layer through which a via-contact is made. These
concepts are still in their early development phases but offer new
possibilities once larger scale production is mastered.
\subsection{CMOS active pixels}
\begin{figure}[h]
\begin{center}
\includegraphics[width=0.45\textwidth]{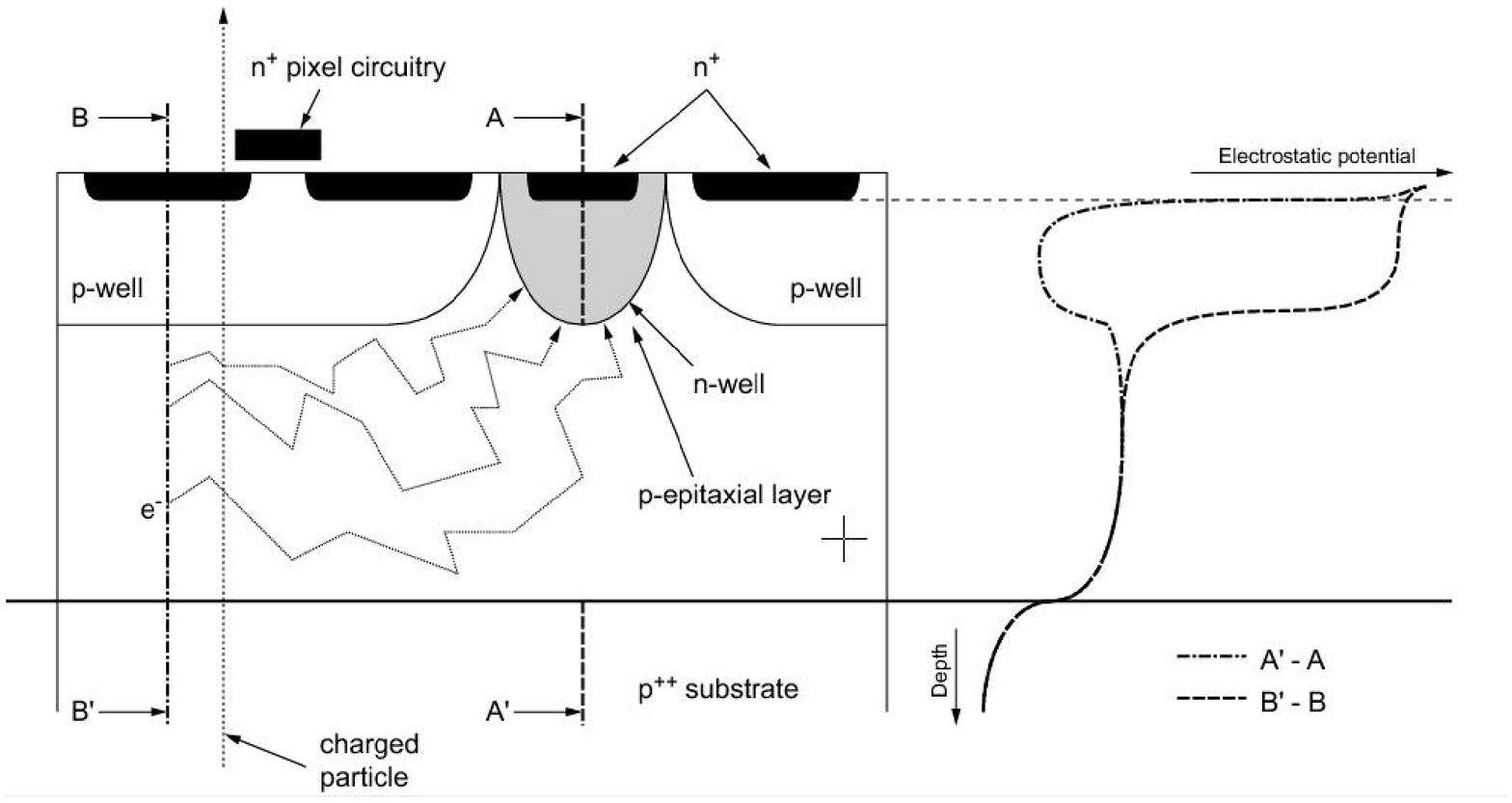}
\vskip 0.2cm
\includegraphics[width=0.35\textwidth]{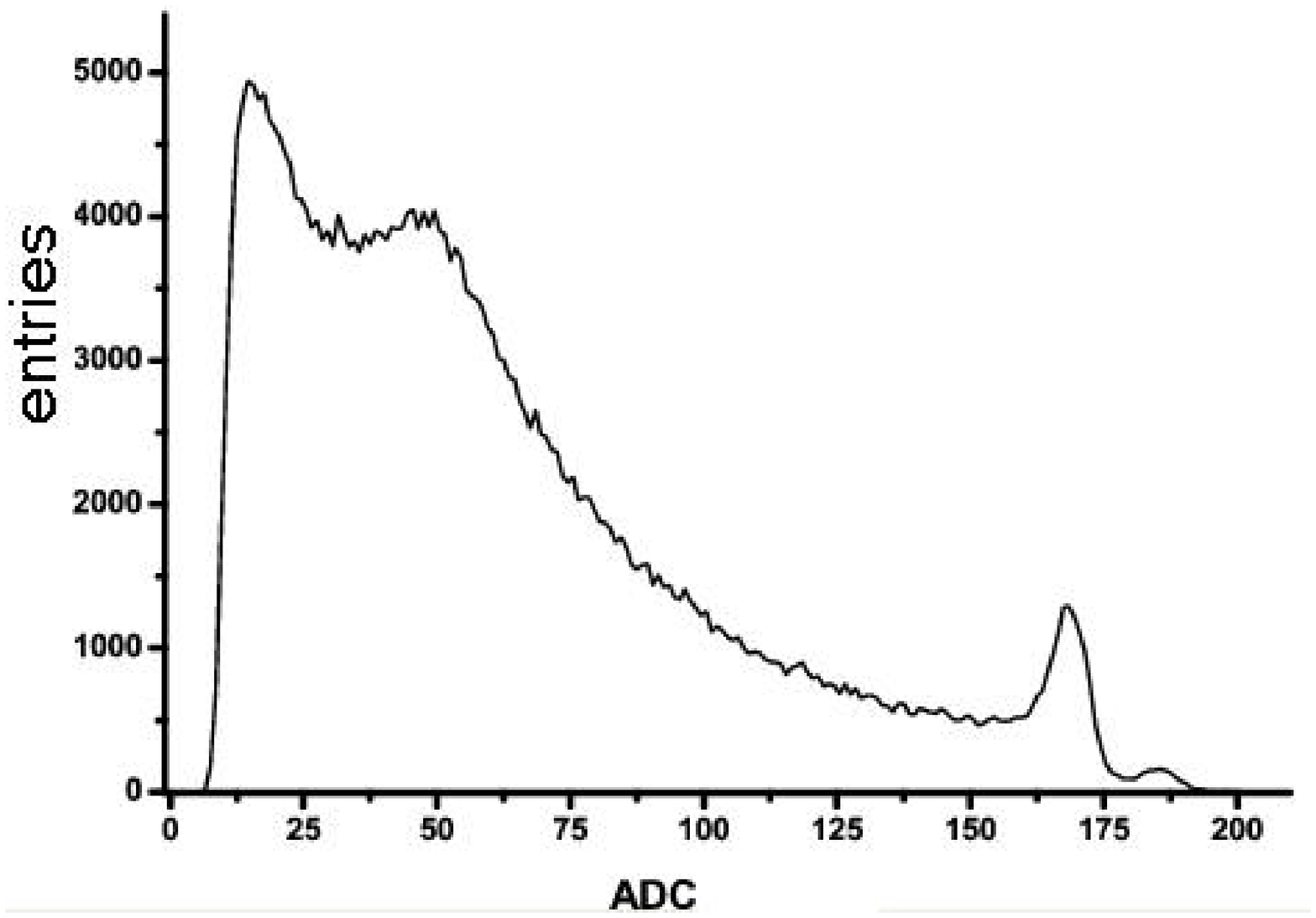}
\end{center}
\caption[]{(top) Principle of a Monolithic Active Pixel Sensor
(MAPS)~\cite{MAPS1}. The charge is generated and collected by
diffusion in the very few $\mu$m thick epitaxial Si-layer.
(bottom) MAPS signal response spectrum to an $^{55}$Fe radioactive
source. The small peak on the right corresponds to full charge
collection.}\label{MAPS}
\end{figure}

\begin{figure}[h]
\begin{center}
\includegraphics[width=0.5\textwidth]{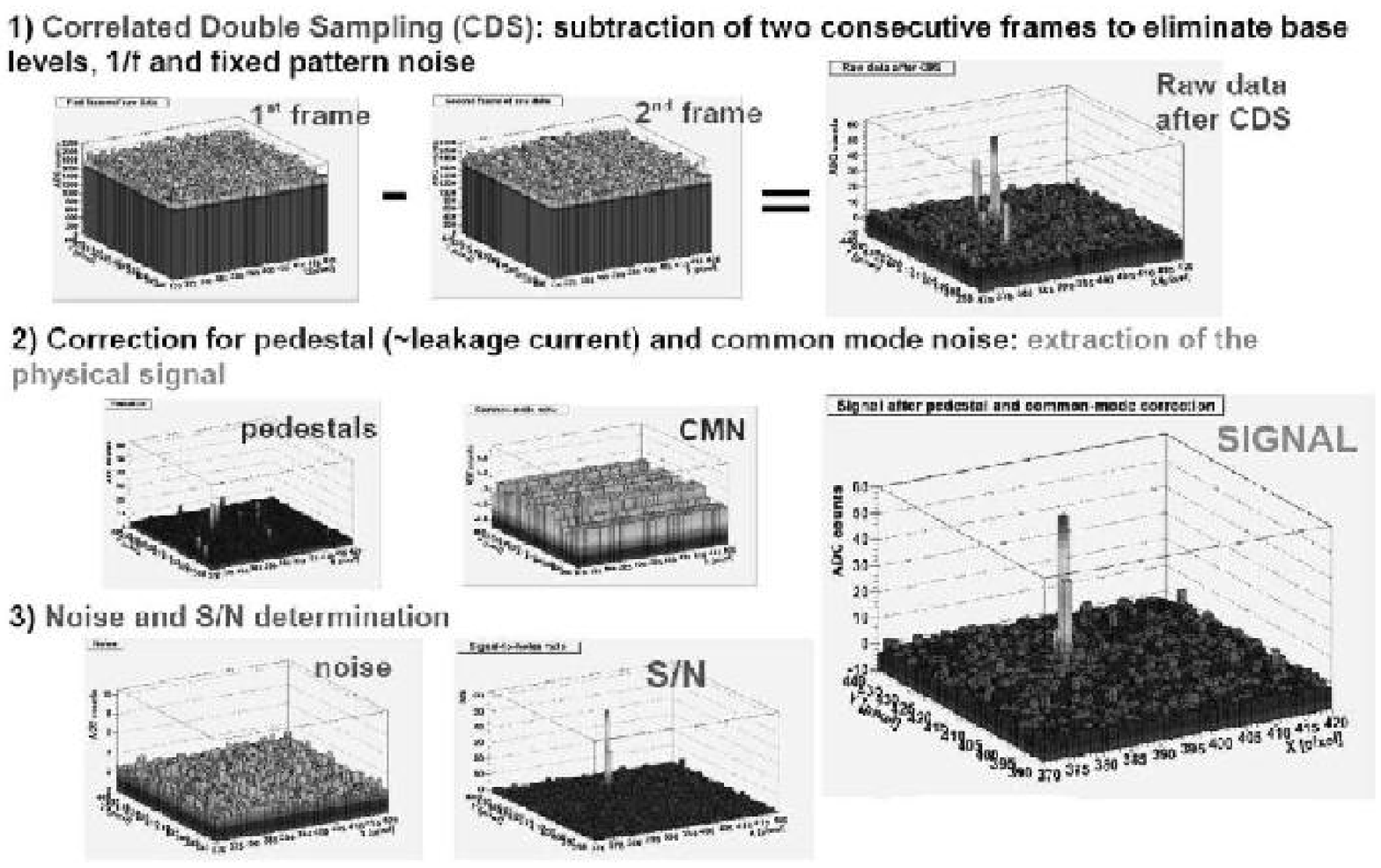}
\end{center}
\caption[]{Readout of a CMOS active pixel matrix (see
text)}\label{MAPS-CDS}
\end{figure}

In some CMOS chip technologies a lightly doped epitaxial silicon
layer of a few to 15$\mu$m thickness between the low resistivity
silicon bulk and the planar processing layer can be used for
charge collection~\cite{meynants98,MAPS1,MAPS2}. The generated
charge is kept in a thin epi-layer atop the low resistivity
silicon bulk by potential wells that develop at the boundary and
reaches an n-well collection diode by thermal diffusion (cf. Fig.
\ref{MAPS}(a)). With small pixel cells collection times in the
order of 100~ns are obtained. The charge collecting epi-layer is
-- technology dependent -- at most 15$\mu$m thick and can also be
completely absent. The attractiveness of active CMOS pixels lies
in the fact that standard CMOS processing techniques are employed
and hence they are potentially very cheap.
CMOS active pixel sensor development are pursued by many groups
which partially collaborate in various projects
(BELLE-upgrade,STAR-upgrade,ILC,CBM at GSI) who use similar
approaches to develop large scale CMOS active pixels, also called
MAPS (Monolithic Active Pixel Sensors)~\cite{MAPS1}. A recent
review can be found in~\cite{winter_PIX2005}. The sensor is
depleted only directly under the n-well diode. The signal charge
is hence very small ($<$1000e) and full charge collection is
obtained only in the depleted region under the n-well electrode
(cf Fig.~\ref{MAPS}(b)). Low noise electronics is therefore the
challenge in this development.
Matrix readout of MAPS is performed using a standard 3-transistor
circuit (line select, source-follower stage, reset) commonly
employed by CMOS matrix devices, but can also include current
amplification and current memory \cite{Dulinski03}. In the active
ares only nMOS transistors are permitted because of the
n-well/p-epi collecting diode which does not permit other n-wells.
For an image two complete frames are subtracted from each other
(correlated double sampling, CDS) to eliminate base levels, 1/f
and fixed pattern noise (see Figure~\ref{MAPS-CDS}). In a second
step pedestals and common mode noise are subtracted to extract the
signal and to determine the remaining noise. Detector sizes up to
19.4$\times$17.4 mm$^2$ with 1M pixels have been tested. The
smallest pixel pitch was 17$\mu$m. Frame speeds of 10$\mu$s for
132x48 pixels have been reached for the BELLE development, with a
noise figure of 30-50e$^-$~\cite{varner_PIX2005}. With other pixel
matrices with slower readout noise values of 15-20~e$^-$, S/N
ratios larger than 20 and spatial resolutions of 1.5$\mu$m
(5$\mu$m) for 20$\mu$m (40$\mu$m) pitch have been
measured~\cite{winter_PIX2005}.
The presently favored technology is the AMS 0.35$\mu$m OPTO
process, which possesses a 10$\mu$m thick epitaxial layer.
Regarding radiation hardness MAPS appear to sustain non-ionizing
radiation (NIEL) to $\sim$10$^{12}$n$_{eq}$. The effects of
ionizing radiation damage (IEL), the main damage source at the
ILC, are threshold shifts and leakage currents in and between nMOS
transistors. The damage effects are less severe when short readout
integration times ($\sim$10$\mu$s) are used. This way doses of
about 10 kGy can be tolerated~\cite{winter_PIX2005}.

The present focus of further development lies in improving the
radiation tolerant design, making 50$\mu$m thin detectors, making
larger area devices for instance by stitching over reticle
boundaries~\cite{AGay03}, and increasing the charge collection
performance in the epi-layer by triple-well~\cite{Rizzo_PIX2005}
or other
techniques~\cite{RAL-Vertex03,Kleinfelder-Portland,LEPSI-Portland}.

%
\subsection{DEPFET pixels}
\begin{figure}[htb]
\begin{center}
\includegraphics[width=0.45\textwidth]{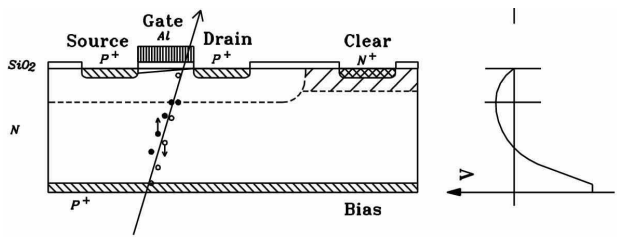}
\end{center}
\caption[]{Principle of operation of a DEPFET pixel structure
based on a sidewards depleted detector substrate material with an
imbedded planar field effect transistor. Cross section (left) of
half a pixel with symmetry axis at the left side, and potential
profile (right).} \label{DEPFET_principle}
\end{figure}

In so-called DEPFET pixel sensors~\cite{kemmer87} FET
transistor is implanted in every pixel on a sidewards
depleted~\cite{gatti84} bulk. Electrons generated by radiation in
the bulk are collected in a potential minimum underneath ($\sim 1
\mu$m) the transistor channel (internal gate) thus modulating its
current (Fig.~\ref{DEPFET_principle}).
Electrons collected in the internal gate are
completely~\cite{sandow_PIX2005} removed by a clear pulse applied
to a dedicated contact outside the transistor. Amplification
values of $\sim$300 pA per collected electron in the internal gate
have been achieved. Further current amplification and storage
enters at the second level stage. The bulk is fully depleted
yielding large signals and the small capacitance of the internal
gate offers low noise operation, for a very large S/N ratio. This
in turn can be used to fabricate very thin devices. Thinning of
pn-diodes to a thickness of 50$\mu$m using a technology based on
wafer bonding and deep anisotropic etching has been successfully
demonstrated~\cite{laci04}.

DEPFET pixels are being developed for three very different
application areas: vertex detection in particle
physics~\cite{DEPFET-TESLA,laci_PIX2005,kohrs_PIX2005}, X-ray
astronomy~\cite{lechner_PIX2005,holl02} and for biomedical
autoradiography~\cite{ulrici05}. With single pixel structures
noise figures below 5e$^-$ and energy resolutions of 131~eV for
6~keV~X-rays have been obtained at room
temperature~\cite{DEPFET_Portland}. The challenges for an ILC
vertex detector are: small pixel cells
($\sim$20$\times$30$\mu$m$^2$), thin, radiation hard sensors
($\sim$50$\mu$m), and fast readout ($\gtrsim$10-20 MHz per matrix
row of 520x4000 pixels). This is pursued by a Bonn-Mannheim-MPI
Munich collaboration.

\begin{figure}[htb]
\begin{center}
\includegraphics[width=0.5\textwidth]{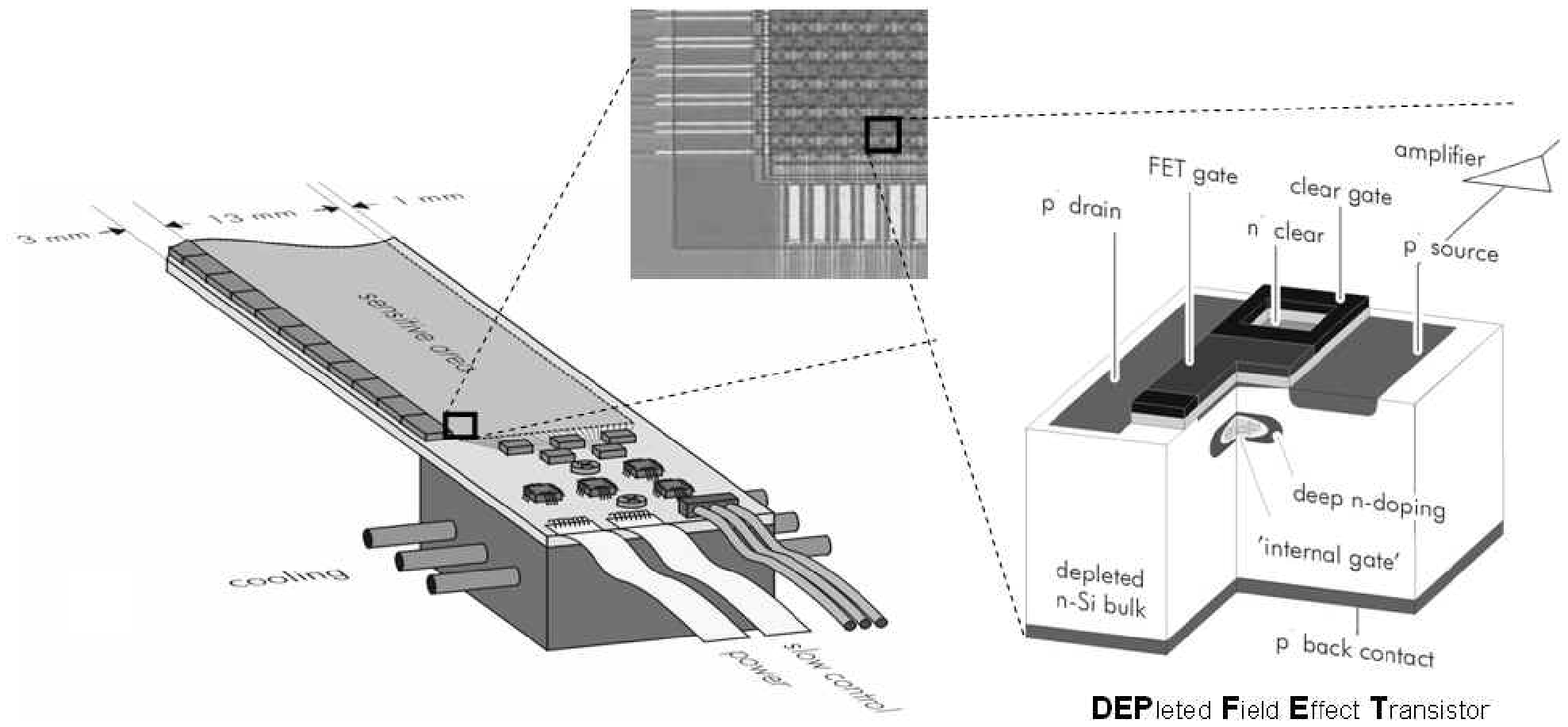}
\vskip 0.5cm
\includegraphics[width=0.48\textwidth]{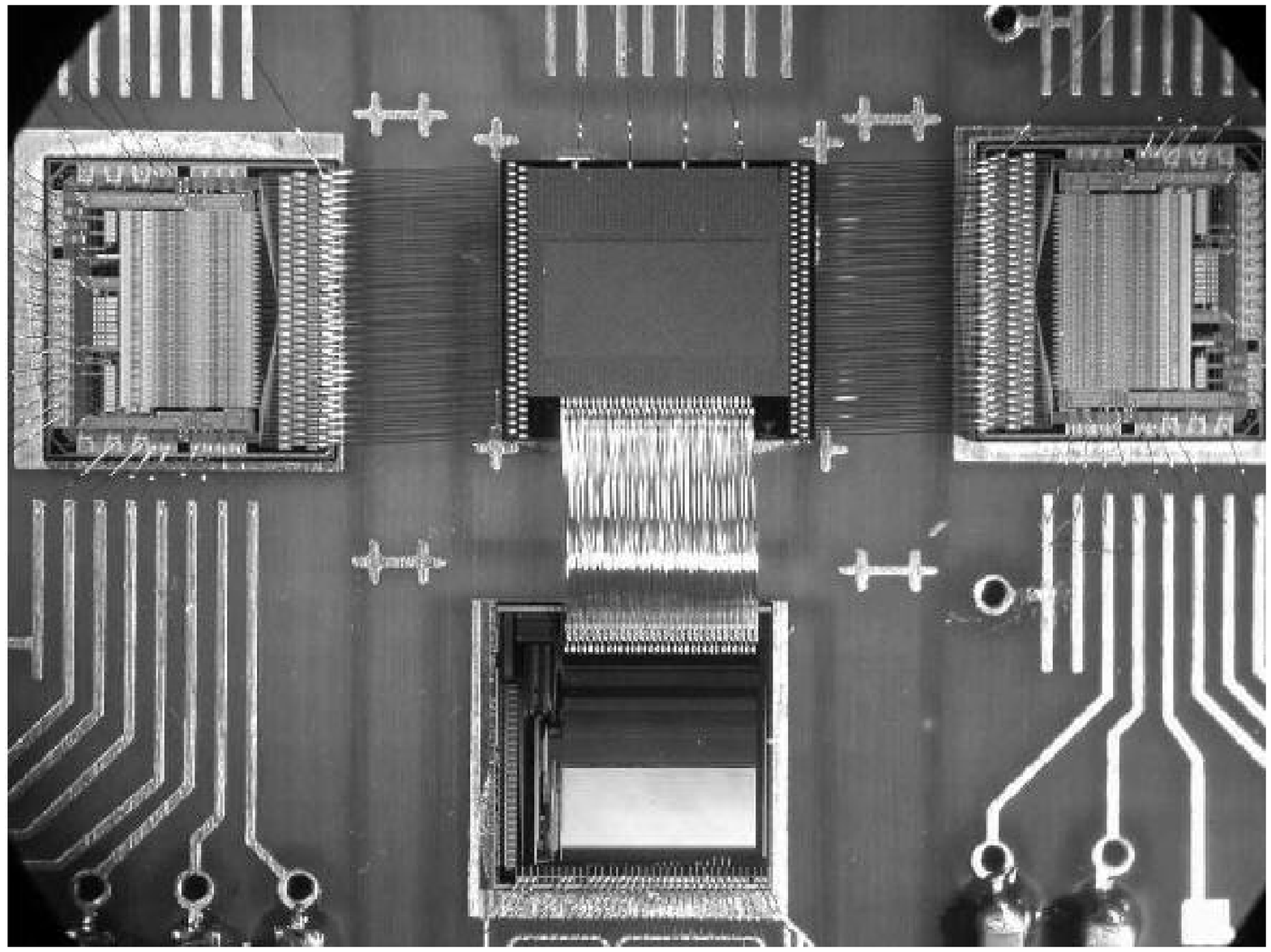}
\end{center}
\caption[]{(top) Sketch of a ILC first layer module with thinned
sensitive area supported by a silicon frame. The enlarged view
show a DEPFET matrix and a DEPFET double pixel structure,
respectively, (bottom) photographs of a DEPFET matrix readout
system (left). The sequencer chips (SWITCHER II) for select and
clear are placed on the sides of the matrix, the current readout
chip (CURO II) at the bottom; (right) stack of the hybrid together
with readout, ADC, and control boards operated in the testbeam.}
\label{DEPFET_TESLA}
\end{figure}

Readout of a DEPFET matrix is done by selecting a row by a gate
voltage from a sequencer chip (SWITCHER) to the external gate. The
drains are connected column-wise delivering their current to a
current-based readout chip (CURO) with amplification and current
storage at the bottom of the
column~\cite{Trimpl02,DEPFET_Portland}. Both chips have been
developed at close to the desired speed for a Linear Collider. A
sketch of a module made of DEPFET sensors is shown in fig.
\ref{DEPFET_TESLA}(top). Figure \ref{DEPFET_TESLA}(bottom) shows a
DEPFET pixel matrix readout system used in the testbeam.

The radiation tolerance, in particular against ionizing radiation,
which is expected to doses of 2 kGy due to beamstrahlung at the
ILC, again is a crucial question. Irradiation with 30 keV X-rays
up to doses of $\sim$10 kGy, about five times the amount expected
at the ILC, have lead to transistor threshold shifts of only about
4 V. Threshold shifts of this order can be coped with by an
adjustment of the corresponding gate voltages supplied by the
SWITCHER chip. The estimated power consumption for a five layer
DEPFET pixel vertex detector at the ILC -- assuming a power duty
cycle of 1:200 -- is only $\sim$5W. Such a performance renders a
very low mass detector without cooling pipes feasible.

\begin{figure}[htb]
\begin{center}
\includegraphics[width=0.5\textwidth]{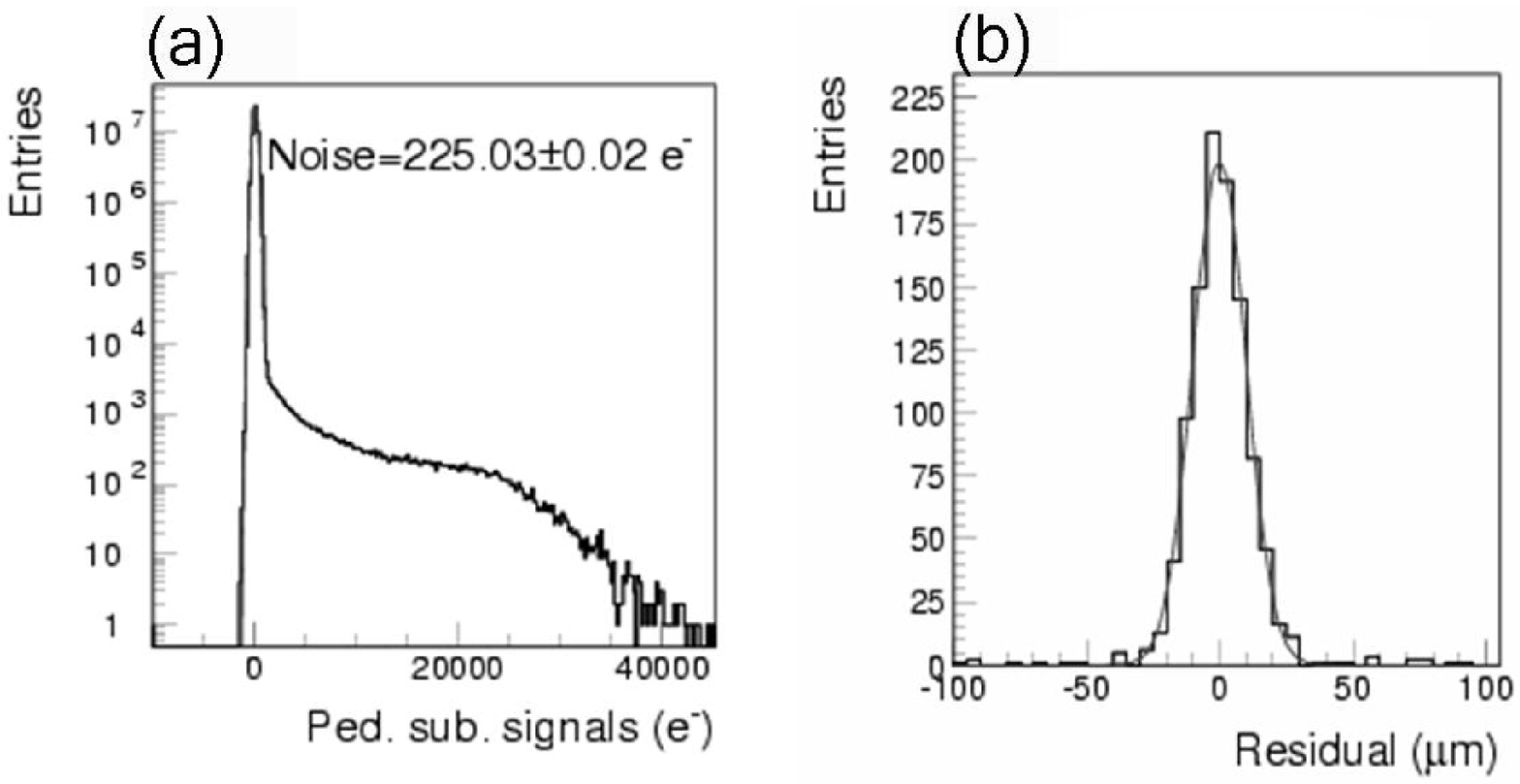}
\end{center}
\caption[]{(a) Raw signal distribution after pedestal subtraction.
(b) Spatial residuals for stiff tracks (multiple scattering
dominated).} \label{TB2}
\end{figure}

A DEPFET pixel matrix with 128$\times$64 pixels has been tested in
a 6 GeV electron beam at DESY. The noise values obtained for the
full system in the test beam including sampling noise of the CURO
chip is 225$e^-$ (see Fig.~\ref{TB2}(a)). The S/N ratio is 144.

\begin{figure}[htb]
\begin{center}
\includegraphics[width=0.5\textwidth]{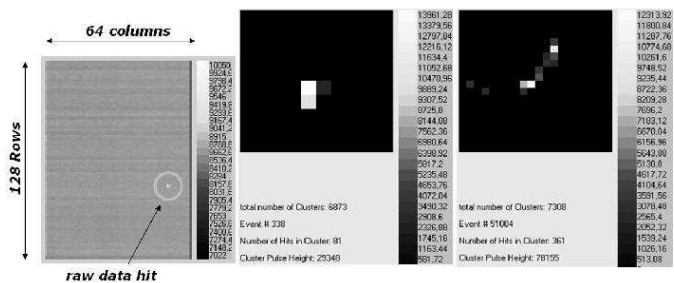}
\end{center}
\caption[]{Raw data event (left) and two typical hit clusters. The
event on the right is consistent with the emission of a
delta-ray.} \label{TB1}
\end{figure}

Figure~\ref{TB1} shows a hit in the DEPFET matrix together with
two typical events with different hit clusters. The event on
Fig.~\ref{TB1}(right) is most likely explained by and consistent
with the emission of a delta-ray which remains in the sensor. This
also qualitatively demonstrates the good reconstruction
capabilities to be expected from a DEPFET vertex detector. At 6
GeV beam energy the spatial residuals are still multiple
scattering dominated. Residuals on the order of 10$\mu$m are
obtained, while with the large S/N value of 144 true space
resolutions in the order of 2$\mu$m should be possible.

\section{Summary}
The large pixel detectors for LHC experiments, based on the hyprid
pixel technology, constitute the state of the art in pixel
detector technology. These detectors are in construction and the
maturity of the technology, including radiation tolerance to 500
kGy doses, has been proven. Immediate spin-off developments are
hybrid pixel detectors with counting capability in which radiation
quanta are individually counted. These developments open up a new
approach to radiological imaging as well as to protein
crystallography with synchrotron radiation. Monolithic or
semi-monolithic detectors, in which detector and readout
ultimately are one entity, are currently being developed in
various forms, largely driven by the needs for particle detection
at the ILC. Most mature at present are CMOS active pixel sensors
using standard commercial technologies on low resistivity bulk,
and DEPFET-pixels, which maintain high bulk resistivity for charge
collection.

\subsubsection*{Acknowledgements}
The author is indebted to all speakers at the PIXEL2005 conference
who provided the most up-to-date material to this review.


\end{document}